\documentstyle[aps,prd,12pt,tighten,amssymb,eqsecnum,axodraw,cite]{revtex}

\title{\vspace*{1.5in}On dynamical mass generation in three dimensional
  supersymmetric U(1) gauge field theory.}

\author{A. Campbell--Smith and N.E. Mavromatos}

\address{Theoretical Physics (University of Oxford), 1 Keble Road,
  OXFORD, OX1 3NP, U.K.,\\and Theory Division, CERN, 1211-CH Geneva 23, Switzerland.}

\newcommand{\slp}{\not p\,}
\newcommand{\slq}{\not \,q\,}

\begin{document}

\maketitle

\vspace*{1cm}
\begin{abstract}
  We investigate and contrast the non--perturbative infra red
  structure of ${\mathcal N}=1$ and ${\mathcal N}=2$ supersymmetric
  non--compact U(1) gauge field theory in three space--time dimensions
  with $N$ matter flavours.  We study the Dyson--Schwinger equations
  in a general gauge using superfield formalism; this ensures that
  supersymmetry is kept manifest, though leads to spurious infra red
  divergences which we have to avoid carefully.  In the ${\mathcal
    N}=1$ case the superfield formalism allows us to choose a vertex
  which satisfies the U(1) Ward identity exactly, and we find the
  expected critical behaviour in the wavefunction renormalization and
  strong evidence for the existence of a gauge independent dynamically
  generated mass, but with no evidence for a critical flavour number.
  We study the ${\mathcal N}=2$ model by dimensional reduction from
  four dimensional ${\mathcal N}=1$ electrodynamics, and we refine the
  old gauge dependence argument that there is no dynamical mass
  generation.  We recognize that the refinement only holds after
  dimensional reduction.
\end{abstract}

\vspace*{-5.5in}
\begin{flushright}
  OUTP-99-23P\\
  CERN-TH/99-111\\
  hep--th/9904173
\end{flushright}
\vspace*{4.5in}



\section{Introduction.}
\label{sec:intro}

Dynamical mass generation is an interesting strong--coupling
phenomenon which requires non--perturbative methods, and has been
studied extensively in $N$--flavour non--compact U(1) gauge theory in
three dimensions (QED${}_3$)
\cite{pisarski84,appelquist86,kondo+nak92,dorey92,maris96,aitch+mav:prb,ijra+gac+mkk+dmcn+nm}.
In this non--supersymmetric model the dynamics drives the coupling to
a non--trivial fixed point in the infra red
\cite{aitch+mav:prb,ijra+gac+mkk+dmcn+nm}: this leads to dynamical
mass generation (and hence chiral symmetry breaking) and critical
behaviour in the normal (chirally symmetric) phase.

The situation for supersymmetric versions of electrodynamics is less
clear.  In an early work on three dimensional ${\mathcal N}=2$
supersymmetric QED, it was argued using component formalism that (at
least in Landau gauge) a dynamical mass is generated
\cite{pisarski84}.  It was pointed out that the vanishing of the
effective potential in supersymmetric field theories meant that the
question of energetic favourability could not be answered simply; it
was conjectured that the issue of the selection of the finite solution
over the vanishing one might be resolved at the level of the effective
action, and the Ward identities arising from the supersymmetry
\cite{pisarski84}.  However, it has been argued on the grounds of
gauge dependence that a (non--perturbative) non--renormalization
theorem in four dimensional ${\mathcal N}=1$ supersymmetric
electrodynamics forbids the dynamical generation of mass
\cite{clark+love}.  On dimensional reduction, four dimensional
${\mathcal N}=1$ supersymmetry reduces to three dimensional ${\mathcal
  N}=2$ extended supersymmetry, and so the non--renormalization
theorem should also hold for the three dimensional extended model.
Recently there has been speculation that the gauge dependence argument
for the four dimensional model is not sound
\cite{kaiselip,appelquist98}.  The analysis of the four dimensional
model employs a simplified form of the U(1) Ward identity, and a
heavily truncated vertex \cite{clark+love}.  This is because in four
dimensions supersymmetry and U(1) gauge invariance require, in
addition to the usual three point vertex, $n$--point vertices for all
$n>3$ (see section \ref{sec:n=2action}).  This problem does not arise
in the three dimensional ${\mathcal N}=1$ model, where supersymmetry
and U(1) gauge invariance require three-- and four--point vertices
only.  Recently there has been evidence from numerical studies
\cite{walker99_1} that the ${\mathcal N}=2$ model does not generate a
mass dynamically, in line with the dimensional reduction of the
non--renormalization theorem, but in contradistinction to the results
of reference \cite{pisarski84}.

An early work on ${\mathcal N}=1$ supersymmetric QED in three
dimensions demonstrated that troublesome logarithmic divergences
present in the wavefunction renormalization of non--supersymmetric
QED${}_3$ are absent in the supersymmetric model, and it was argued
that this allowed dynamical mass generation to take place for numbers
of flavours less than a critical number \cite{koopmans89}, which was
related to the critical flavour number of non--supersymmetric
QED${}_3$ \cite{appelquist86}.  The analysis was done in component
formalism, in Landau gauge and with a trivial three--point vertex (set
to unity) which does not satisfy the U(1) Ward identity except in the
limits of zero transferred momentum and trivial (unit) wavefunction
renormalization.

It is therefore of interest to look more closely at the infra red
structure of both ${\mathcal N}=1$ and ${\mathcal N}=2$ supersymmetric
U(1) gauge field theory.  The actual dynamically generated mass is the
$p\rightarrow 0$ limit of the mass function and is the pole of the
matter two point correlation function; as such, it must be a gauge
invariant object.  Only if the $p\rightarrow 0$ limit leaves a gauge
independent constant can it be said with certainty that a mass has
been dynamically generated.  It is therefore crucial to give a careful
treatment of gauge dependence, and to look closely at the full vertex.
Since supersymmetry cannot be dynamically broken in a supersymmetric
U(1) gauge field theory \cite{witten82}, we do not have to consider
the possibility of different dynamical masses for each component of
the multiplet.  With these issues in mind, we choose to work in
superfield formalism, which keeps supersymmetry manifest, and we will,
as far as possible, work in a general gauge.  In components
computations (see {e.g.} reference \cite{walker99_2}) it is necessary
to use the U(1) Ward identity and the Ward identities arising from
supersymmetry to constrain the vertices: the advantage of the
superfield formalism is that we only need to examine the U(1) Ward
identity, for the formalism ensures that the supersymmetry Ward
identities are satisfied.  In particular, by using the superfield
formalism we will be able to choose a three--point vertex which
satisfies the U(1) Ward identity exactly in the ${\mathcal N}=1$ case.
It is known \cite{clark77,clark+love} that superfield formalism for
gauge theories results in spurious infra red divergences arising from
the propagation of a gauge artifact: by careful choice of
approximations and, where necessary, gauge, we can avoid most of the
infra red divergences and they do not spoil our results.

We use the same approach to study the non--perturbative physics of
both the ${\mathcal N}=1$ and ${\mathcal N}=2$ models:  we construct a
non--perturbative {\sl ansatz} for the matter superfield propagator,
and study the behaviour of the functions appearing therein through the
Dyson--Schwinger equations.  In principle there are Dyson--Schwinger
equations for the matter superfields, the gauge superfield, and for
the full vertex.  The gauge superfield Dyson--Schwinger equation will be
resummed to leading order in $1/N$, where $N$ is the number of matter
flavours; this is exactly what is done in normal QED${}_3$.  The
vertex will be constrained by the U(1) Ward identity.  When the
results of the resummation and a choice for the vertex are put into
the Dyson--Schwinger equation for the matter superfield, coupled
integral equations for the self energy and wavefunction
renormalization result, which have to be solved.  As with normal
QED${}_3$, approximations have to be sought in order to look for
solutions of the equations.

For the ${\mathcal N}=1$ model we will present two approaches for
studying the Dyson--Schwinger equations.  First, we consider some of
the simplest approximations, which are sufficient when applied to
normal QED${}_3$ to demonstrate the existence of dynamical mass
generation (at least in Landau gauge) \cite{kondo+nak92}.  The
advantages of making these computationally convenient approximations
are twofold: they allow us to work in a general gauge throughout, and
admit conversion of the integral equation to an equivalent
differential equation.  The differential equation is simple to solve,
and has a gauge invariant, constant solution in the $p\rightarrow 0$
limit.  The second method incorporates the full vertex, consistent
with the U(1) Ward identity.  Ideally we would like to probe the gauge
dependence in exactly the same way again; however, the richer
structure of the integral equations arising from this approach only
admit more direct methods.  In both approaches we find the expected
critical behaviour in the wavefunction renormalization
\cite{appelquist81,aitch+mav+mcneill,kondo+mura97}
\begin{equation}
  \label{expectedwf}
Z(p) \sim \left( \frac{p}{\alpha} \right)^\gamma , \qquad \gamma \sim
{\mathcal O}(1/N),
\end{equation}
and exhibit the possibility of a finite dynamically generated mass.
In contrast to reference \cite{koopmans89} we find no evidence for a
critical flavour number, above which there is no dynamical mass
generation.

For the ${\mathcal N}=2$ model we again make some simple
approximations which allow us to probe the gauge dependence.  Again we
find the expected critical behaviour in the wavefunction
renormalization (\ref{expectedwf}), but this time we find that there
is no gauge independent dynamically generated mass.  This is based on
a refinement of the gauge dependence argument of reference
\cite{clark+love}; we note that the refinement relies on the existence
of the scale arising from the compactified dimension, so that the
non--renormalization theorem for the four dimensional model
\cite{clark+love} may still be evaded \cite{kaiselip,appelquist98}.

In section \ref{sec:action} we construct the action functional for a
${\mathcal N}=1$ supersymmetric theory of $N$ matter flavours
interacting with a U(1) gauge field in superfield formalism.  We also
give the dressed propagators for the model which result from this
action.  We construct the U(1) Ward identity in section \ref{sec:brs}
using the BRS method, and in section \ref{sec:ds} we construct the
Dyson--Schwinger equations for the model, which we use to analyse the
non--perturbative properties of the theory.  We present and briefly
discuss our simple computation in section \ref{sec:gauge}, in which we
are able to probe the gauge dependence of the mass function by making
convenient approximations in the integral equations.  Our more
complete computation follows in section \ref{sec:fullcomp}, where we
choose a vertex which satisfies the U(1) Ward identity.  Here we
cannot use the elegant methods of section \ref{sec:gauge}, but we are
able to demonstrate that the solution obtained there persists in this
more complete computation.  To conclude this part of the paper, we
also discuss the question of whether the finite mass solution is
preferred to the vanishing mass solution, in the context of the
suggestion \cite{pisarski84} that this can be answered by appealing to
the effective action and the supersymmetry Ward identities.  We find
that no extra information can be obtained in this way, in contrast to
the models of references \cite{alvarez78,ciuchini95,diamandis98} which,
crucially, have extra constraints.

We turn to the ${\mathcal N}=2$ extended model in section
\ref{sec:n=2action}, where we build an action functional and dressed
propagators for four dimensional ${\mathcal N}=1$ supersymmetric
$N$--flavour electrodynamics.  We construct the Dyson--Schwinger
equations for the model in section \ref{sec:n=2ds} and then compactify
one dimension to obtain the coupled integral equations for the three
dimensional ${\mathcal N}=2$ model we wish to analyse.  We compute the
wavefunction renormalization and refine the argument of reference
\cite{clark+love} for the non--existence of dynamical mass generation.

We append discussion and concluding remarks in section
\ref{sec:discuss}, and briefly contrast with the situation in
non--supersymmetric QED${}_3$.  In the appendices we collect some
useful results and basic features of ${\mathcal N}=1$ supersymmetry in
three and four dimensions, which are essential to the analysis of the
models we consider.

\section{The ${\mathcal N}=1$ Action.}
\label{sec:action}

The ${\mathcal N}=1$ rigid superspace in three dimensions can be
parameterized by the usual three space--time coordinates and two
Grassmann--odd coordinates arranged in the Majorana spinor
$\theta^\alpha$, where $\alpha \in \{ 1,2 \}$ \cite{gates83:super},
which we denote collectively by the symbol $z$.  We collect some
basic features of this ${\mathcal N}=1$ superspace and superfields in
appendix \ref{app:3d}.  Consider complex scalar ``matter''
superfields $\Phi$, $\Phi^*$ (mappings from superspace to $\mathbb C$)
which transform under local U(1) transformations in the familiar way:
\begin{eqnarray}
\Phi(z) \longrightarrow &\Phi^\prime (z^\prime)& = e^{+ieG(z)} \,\Phi(z),
\nonumber\\
\Phi^* (z) \longrightarrow &\Phi^*{}^\prime (z^\prime)& = \Phi^* (z)\,
e^{-ieG(z)} .
\end{eqnarray}
Here $e$ is the (dimensionful) gauge coupling and $G(z)$ must be a
real scalar superfield (a mapping from superspace to $\mathbb R$) to
preserve the superfield nature of $\Phi^\prime$; in order to build an
action functional invariant under local U(1) transformations, we must
construct covariant derivatives which transform in the same way as the
matter superfields themselves:
\begin{equation}
\nabla_\alpha \longrightarrow \nabla^\prime_\alpha = e^{ieG(z)}
\nabla_\alpha e^{-ieG(z)} .
\end{equation}

The covariant derivatives can be written in terms of a real spinor
(superfield) connexion $\Gamma_\alpha$ as follows:
\begin{equation}
\nabla_\alpha = D_\alpha - i e \Gamma_\alpha ;
\end{equation}
here $D_\alpha$ is the normal spinorial derivative, covariant with
respect to supersymmetry transformations \cite{gates83:super}.  The
connexion transforms in the usual way under infinitesimal local U(1)
transformations:
\begin{equation}
\Gamma_\alpha (z) \longrightarrow \Gamma^\prime_\alpha (z^\prime) =
\Gamma_\alpha (z) + D_\alpha G(z) .
\end{equation}

We wish to consider a model with ${\mathcal N}=1$ supersymmetry, local
U(1) gauge invariance and $N$ matter flavours.  The required action
then comprises three parts: the gauge invariant classical field
strength term for the connexion $\Gamma_\alpha$, a (Lorentz) gauge
fixing term and a locally U(1) invariant kinetic term for the matter
superfields $\Phi$ and $\Phi^*$:
\begin{equation} \label{n=1action}
S = S_{g}^{\mathrm{class}} + S_{g}^{\mathrm{GF}} + S_m ;
\end{equation}
\begin{eqnarray}
S_g^{\mathrm{class}} &=& \int d^3 x\, d^2\theta \; \Gamma_\alpha \left(
-\frac{1}{8} D^\eta D^\alpha D^\beta D_\eta \right) \Gamma_\beta
,\nonumber\\
S_g^{\mathrm{GF}} &=& \int d^3 x\, d^2 \theta\; \Gamma_\alpha \left(
\frac{1}{4\xi} D^\alpha D^2 D^\beta \right) \Gamma_\beta ,\nonumber\\
S_m &=& \int d^3 x\, d^2\theta\; \left(-\frac{1}{2}\right) \left[
\nabla^\alpha \Phi \right]^* \left[ \nabla_\alpha \Phi \right] .
\end{eqnarray}
We have included in the matter part an implicit sum over $N$ flavours,
which do not interact with each other but interact with the same gauge
field.  The parameter $\xi$ is the familiar gauge parameter;
supersymmetric Feynman and Landau gauges are given by $\xi=1,0$
respectively.  From this action it is easy to derive the renormalized
propagators for the matter fields and connexion:
\begin{eqnarray} \label{props}
\Delta(p;12) &=& i \frac{ Z(p) D^2 -
\Sigma(p)}{Z^2(p)\,p^2+\Sigma^2(p)} \delta^2(12), \nonumber\\
\Delta_{\alpha\beta} (p;12) &=& -i
\frac{1}{p^4}\frac{1}{1+\alpha/|p|} \left[ (1\!+\!\xi) p_{\alpha\beta}
\, D^2 - (1\!-\!\xi) C_{\alpha\beta} \, p^2 \right] \delta^2(12).
\end{eqnarray}
The gauge field propagator includes the effects of massless matter
loops (to leading order in $1/N$) in the vacuum polarization factor
\cite{pisarski84} \[(1+\alpha/|p|)^{-1}.\] Both the spinor and scalar
components of the matter superfield $\Phi$ contribute to this
correction, and so the vacuum polarization is exactly twice that of
non--supersymmetric U(1) gauge theory \cite{pisarski84,koopmans89},
and the effective ultra violet scale $\alpha$ is given by
\begin{equation}
  \alpha \doteq \frac{e^2 N}{4}.
\end{equation}
The second rank antisymmetric symbol $C_{\alpha\beta}$ appearing in
the gauge field propagator acts as a metric for spinor indices
\cite{gates83:super}, (see appendix \ref{app:3d}) and we have written
three--vectors in a convenient spinor notation: $p_{\alpha\beta}$ is a
symmetric second rank spinor.  In all the above an abbreviated
superspace notation has been used where, for instance,
\begin{equation}
  \delta^2(12) \equiv \delta^2(\theta_1 - \theta_2) \doteq - \left
  ( \theta_1 - \theta_2\right)^2.
\end{equation}
The interaction piece of the action is given by:
\begin{equation}
S_{\mathrm{INT}} = \int d^3 x\, d^2\theta\; \left( -\frac{1}{2}
C^{\alpha\beta} ie \, \Gamma_\alpha \Phi D_\beta \Phi^* + \frac{1}{2}
C^{\alpha\beta} ie \, \Gamma_\alpha \Phi^* D_\beta \Phi - \frac{1}{2}
e^2 C^{\alpha\beta} \Gamma_\beta \Gamma_\alpha \Phi^* \Phi \right).
\end{equation}
The Feynman rules derived from this action are given at the end of
appendix \ref{app:3d}.  Finally, note that we have not used any
Wess--Zumino type gauge fixing for the connexion superfield; this is
not a gauge invariant truncation (nor one which respects
supersymmetry), and we wish, as far as is possible, to investigate the
full gauge dependence of the non--perturbative correlation functions.

\section{BRS Invariance And Ward Identities.}
\label{sec:brs}

Following the BRS approach, to compute the Ward identities for this
model we promote the gauge fixed action (\ref{n=1action}) to an action
invariant under an enhanced gauge symmetry.  The gauge fixing term in
(\ref{n=1action}) transforms under infinitesimal gauge transformations
as follows:
\begin{equation}
\delta_{\mathrm G} \frac{1}{4\xi} \Gamma_\alpha \left( D^\alpha D^2 D^\beta \right)
\Gamma_\beta = \frac{1}{\xi} G \left( \square D^\alpha \Gamma_\alpha
\right).
\end{equation}
To make a BRS invariant action, we add a ghost term to the action,
whose transformation properties cancel those of the gauge fixing term:
\begin{equation}
S^{\mathrm{ghost}} = \int d^3 x \, d^2 \theta\; \left(- u \square w \right),
\end{equation}
where $u$, $w$ are real Grassmann--odd scalar ghost superfields.  The
full action is then invariant under the extended gauge symmetry
\begin{eqnarray}
\delta_\eta \Phi &=& ie \left(\eta w\right) \, \Phi, \nonumber\\ 
\delta_\eta \Phi^* &=& -ie \Phi^* \, \left(\eta w\right), \nonumber\\
\delta_\eta \Gamma_\alpha &=& D_\alpha \left( \eta
w\right),\nonumber\\
\delta_\eta u &=& -\frac{1}{\xi} \eta \, D^\alpha \Gamma_\alpha ,
\nonumber\\
\delta_\eta w &=& 0; 
\end{eqnarray}
here $\eta$ is a real Grassmann--odd number.  Note that $\delta^2_\eta
=0$.  The generating functional can now be written
\begin{equation}
{\mathcal Z} = \int I\!\!D \Phi \, I\!\!D \Phi^* \, I\!\!D \Gamma_\alpha \,
I\!\!D u \, I\!\!D w \; e^{iS + iS_{\mathrm{s}} -i\Delta_{\mathrm{s}}} ,
\end{equation}
where
\begin{eqnarray}
S_{\mathrm{s}} &=& \int d^3 x\, d^2\theta\; \left( J^*\Phi
+ J\Phi^* +K^\alpha \Gamma_\alpha +\sigma u + \tau w \right),
\nonumber\\
\Delta_{\mathrm{s}} &=& \int d^3 x\, d^2\theta\; \left(
J^* \delta_\eta \Phi + J \delta_\eta \Phi^* +K^\alpha \delta_\eta
\Gamma_\alpha +\sigma \delta_\eta u + \tau \delta_\eta w \right).
\end{eqnarray}
The set $\{J,J^*,K^\alpha,\sigma,\tau\}$ are superfield sources; the
last three are Grassmann--odd valued and $K$ is a spinor.  Using the
fact that $\eta^2 =0$, the Ward identity immediately follows
\begin{eqnarray}
  \label{wione}
0 &=& \int I\!\!D \Phi \, I\!\!D \Phi^* \, I\!\!D \Gamma_\alpha \,
I\!\!D u \, I\!\!D w \; e^{iS +iS_{\mathrm{s}}} \times \nonumber\\
&& \qquad \times \left( \int d^3 x\, d^2\theta\; \left( J^* \delta_\eta \Phi
    + J \delta_\eta \Phi^* +K^\alpha \delta_\eta \Gamma_\alpha +\sigma
    \delta_\eta u + \tau \delta_\eta w \right) \right).
\end{eqnarray}
To interpret this in terms of correlation functions for the model, we
construct the quantum effective action (the index $c$ indicates a
classical value):
\begin{eqnarray}
\Gamma\left[ \Phi_c,\Phi^*_c,\Gamma_\alpha^c,u_c,w_c \right] &=& -i \ln
{\mathcal Z}\left[J,J^*,K^\alpha,\sigma,\tau\right] \nonumber\\ 
&& \qquad - \int d^3 x\, d^2\theta\; \left( J\Phi_c^* + J^* \Phi_c +
  K^\alpha \Gamma_\alpha^c +\sigma u_c + \tau w_c \right);
\end{eqnarray}
the sources are given by functional derivatives of the effective action
\begin{eqnarray}
  &&\frac{\delta \Gamma}{\delta\Phi_c} = - J^*, \qquad \frac{\delta
  \Gamma}{\delta\Phi_c^*} = - J, \nonumber\\ 
  &&\frac{\delta \Gamma}{\delta\Gamma_\alpha^c} = K^\alpha,  \nonumber\\
  &&\frac{\delta \Gamma}{\delta u_c} = \sigma, \qquad\quad \frac{\delta
  \Gamma}{\delta w_c} = \tau .
\end{eqnarray}

\begin{eqnarray}
\Gamma\left[ \Phi_c,\Phi^*_c,\Gamma_\alpha^c,u_c,w_c \right] &=& \int
d^5 x\, d^5 y \; \left( \Phi^*_c (x) \Delta^{-1} (x-y) \Phi_c (y) +
\Gamma_\alpha (x) \left( \Delta^{-1} (x-y) \right)^{\alpha\beta}
\Gamma_\beta (y) \right) \nonumber\\
&+& \int d^5 x \,d^5 y \,d^5 z \; \left( \Phi^*_c (x) \,
\Gamma_\alpha^c (y) \, \frac{e}{2}G(x,y,z) D^\alpha \Phi_c (z) \right.\nonumber\\
&& \qquad\qquad + \left. \Phi_c (x) \,
\Gamma_\alpha^c (y) \, \frac{e}{2}G(x,y,z) D^\alpha \Phi^*_c (z) \right) +
\cdots
\end{eqnarray}
Now we can carry out the functional integral in the identity
(\ref{wione}) to obtain
\begin{equation}
0= \int d^5 x \left( ie \frac{\delta\Gamma}{\delta \Phi^*_c} \Phi^*_c
w_c - ie \frac{\delta\Gamma}{\delta \Phi_c} w_c \Phi_c + D_\alpha
\frac{\delta\Gamma}{\delta\Gamma_\alpha^c} w_c + \frac{1}{\xi}
\frac{\delta\Gamma}{\delta u_c} D^\alpha \Gamma_\alpha^c \right).
\end{equation}
Taking functional
derivatives with respect to $w_c$, $\Phi^*_c$ and $\Phi_c$ we obtain
the Ward identity in configuration space:
\begin{equation}\label{wiconfig}
0= D^2(y) G(x,y,z) + i \Delta^{-1} (x-z) \delta(x-y) -i \Delta^{-1}
(x-z) \delta(z-y),
\end{equation}
which in momentum space reads
\begin{equation}\label{wimom}
D^2(p-q) G(p,p-q,q) = i \Delta^{-1} (p) -i \Delta^{-1} (q) .
\end{equation}
The advantage of using a superspace formulation is now apparent, for
using the identity
\begin{equation}
\left(D^2 (p) \right)^2 = -p^2,
\end{equation}
the Ward identity (\ref{wimom}) can be inverted.  In fact, as we shall
see, the structure of the Dyson--Schwinger equations is such that it
can always be organized so that $D^2(p-q)$ acts to the left of the
full vertex, and so the Ward identity can be used exactly as it
appears in (\ref{wimom}).  Of course, as with the non-supersymmetric
U(1) theory, the Ward identity does not constrain the transverse part of the
vertex, for which
\begin{equation}
D^2(p-q) G(p,p-q,q) = 0.
\end{equation}
Since, as discussed above, the integral equations can be arranged so
that the full vertex only appears with an accompanying $D^2$, this
will not be a problem for our analysis.  Moreover, in not having to
invert the Ward identity, we will not introduce any extra kinematical
singularities as $p\rightarrow q$; this is known to cause problems in
the non--supersymmetric model \cite{curtis90}.

In principle the four--point vertex could also be constrained by the
U(1) Ward identity; however, the form depends in a complicated way on
the three--point vertex and it is sufficient for us to choose a
trivial function for our four--point vertex.  As we will show, the
graph which includes the four--point vertex contributes only to the
wavefunction renormalization, and the contribution vanishes in
supersymmetric Feynman gauge.
 
\section{The Dyson--Schwinger Equations.}
\label{sec:ds}

The non--perturbative properties of the model are determined through
the Dyson--Schwinger equations.  These equations lead to coupled
integral equations for the self energy and wavefunction
renormalization, which can be solved (in principle) to yield the
dynamically generated mass:
\begin{equation}
  \label{dynmassdef}
  M\doteq \lim_{p\rightarrow 0} M(p) \equiv \lim_{p\rightarrow 0}
  \frac{\Sigma(p)}{Z(p)}.
\end{equation}

\begin{figure}
\begin{center}
\begin{picture}(260,80)(0,0)
\Line(0,30)(60,30)
\GCirc(30,30){7}{0.5}
\LongArrow(20,20)(40,20)
\Text(30,10)[]{$p$}
\Text(5,40)[]{$\Phi$}
\Text(55,40)[]{$\Phi^*$}
\Text(70,30)[]{$=$}
\Line(80,30)(140,30)
\PhotonArc(110,48)(15,0,360){3.5}{7}
\GCirc(110,63){5}{0.5}
\LongArrow(100,20)(120,20)
\Vertex(110,30){4}
\LongArrowArcn(110,44)(12,130,50)
\Text(110,48)[]{$q$}
\Text(110,10)[]{$p$}
\Text(85,40)[]{$\Phi$}
\Text(135,40)[]{$\Phi^*$}
\Text(150,30)[]{$-$}
\Line(160,30)(260,30)
\GCirc(210,30){7}{0.5}
\PhotonArc(210,30)(25,0,180){4}{5.5}
\GCirc(210,55){5}{0.5}
\LongArrowArcn(210,38)(25,130,50)
\Text(212,72)[]{$p-q$}
\Vertex(185,30){4}
\Vertex(235,30){1}
\Text(165,40)[]{$\Phi$}
\Text(255,40)[]{$\Phi^*$}
\LongArrow(200,20)(220,20)
\Text(210,10)[]{$q$}
\end{picture}
\caption{\label{fig:dseq} Schematic form of the Dyson--Schwinger
  equation for the non--perturbative matter two point correlation
  function.  Solid lines represent matter superfield propagators,
  and wavy lines gauge superfield propagators; blobs indicate full
  non--perturbative quantities.}
\end{center}
\end{figure}
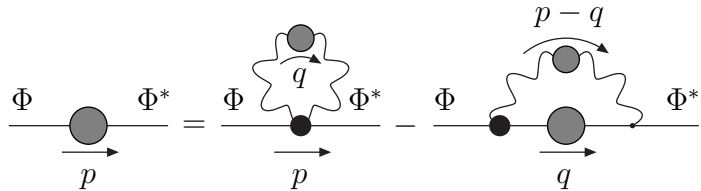

The truncated Dyson--Schwinger equation we will use is shown
graphically in figure \ref{fig:dseq}.  The graph on the left hand side
is a convenient shorthand for the difference between the full inverse
propagator and the bare inverse propagator.

The graph on the left hand side of figure \ref{fig:dseq} is given by
\begin{eqnarray} \label{dslhs}
  &&-i \int d^2\theta_1\,d^2\theta_2 \; \Phi(-p,\theta_1) \,\left[ Z(p)
   D^2 (p) + \Sigma(p) \right] \,\delta^2(12) \,\Phi^* (p,\theta_2)
   \nonumber\\
  &&+i \int d^2\theta \; \Phi(-p,\theta) \, D^2(p) \,
   \Phi^*(p,\theta)\nonumber\\ 
  &&\quad = -i (Z(p)-1) \int d^2\theta \; \Phi(-p,\theta) D^2 (p) \Phi^*(p,\theta)
   -i \Sigma (p) \int d^2\theta\; \Phi(-p,\theta) \Phi^*(p,\theta).
\end{eqnarray}
The first ``seagull'' graph on the right is simply
\begin{equation} \label{seagull}
  \int \frac{d^3q}{(2\pi)^3} \; d^2\theta_1 \, d^2
  \theta_2 \; \Phi(-p,\theta_1) \left[ -\frac{e^2}{2} G_4 C^{\beta\alpha}
  \delta^2(12) \Delta_{\alpha\beta}(P;12) \right] \Phi^* (p,\theta_2), 
\end{equation}
and the last graph is given by $(P \doteq p-q)$
\begin{equation} \label{maingraph}
  \int \frac{d^3q}{(2\pi)^3} \; d^2\theta_1 \, d^2 \theta_2 \;
  \Phi(-p,\theta_1) \, \left[ -\frac{e^2}{4} G_3(p,P,q) D^\alpha (q)
  D^\beta (q) \Delta (q;12) \right] \Delta_{\alpha\beta} (P;12) \, \Phi^*(p,\theta_2).
\end{equation}
Our use of a symmetric second rank spinor notation for three vectors
immediately implies the following identities
\begin{eqnarray}
  \label{SuSpIds}
&&C^{\alpha\beta} \,p_{\alpha\beta} = 0,\nonumber\\
&&q^{\alpha\mu}\,p_{\mu\beta} = \delta^\alpha_\beta \,p\cdot q.
\end{eqnarray}
Note that the second of these identities indicates that the $C$
symbol from the four--point vertex in the last graph projects out only
the $(1\!-\!\xi)$ component of the gauge field propagator, and indeed
the entire graph would vanish in the supersymmetric Feynman gauge,
$\xi=1$.

Upon substitution of the propagators (\ref{props}), the superspace
integrations can be performed using the identities above along with
the relations (\ref{SpIds}) and (\ref{DIds}) in appendix \ref{app:3d}.  We will
present two computations in what follows: first, a computation in
which the simplest approximations are made: $G_3(p,P,q)$ is set
to $Z(q)$, and the bifurcation method is used, in which $\Sigma$ is
set to vanish in the denominators of all the kernels.  These
approximations are drastic but computationally extremely convenient,
and are sufficient to demonstrate the existence of a dynamically
generated mass in non--supersymmetric U(1) gauge theory
\cite{kondo+nak92}.  These approximations will give us enough
flexibility to probe the gauge dependence of the solution.  Second, we
present a more complete computation, in which the full three--vertex
consistent with the Ward identity (\ref{wimom}) is used.

The treatments in the two cases follow a similar strategy: inserting
the propagators (\ref{props}) into equations (\ref{maingraph}) and
(\ref{seagull}) along with a choice for the three--point vertex, the
superspace parts of the integrals can be unpacked and reassembled
using the identities (\ref{SuSpIds}), (\ref{SpIds}) and (\ref{DIds})
to appear as functions of $\{Z,\Sigma,p,q\}$ multiplying the two
superspace structures:
\begin{eqnarray}
  \label{SuSpStrs}
  &&\int d^2\theta\; \Phi(-p,\theta) \, D^2(p) \, \Phi^* (p,\theta)
  ;\nonumber\\
  &&\int d^2\theta\; \Phi(-p,\theta) \, \Phi^* (p,\theta).
\end{eqnarray}
Comparison with equation (\ref{dslhs}) shows that the function
multiplying the first of these structures is to be identified with the
wavefunction renormalization, and that multiplying the second with the
self energy function.  From this point the usual procedure followed
with the non--supersymmetric model can be adopted to study the coupled
integral equations which multiply the structures above.  The angular
integrations can be performed, leaving integral equations now over the
variable $|q|$ only, which can in principle be evaluated.  In
practice, the kernels have to be approximated with low--momentum
expansions.  This is justified, for the (supersymmetric and
non--supersymmetric) model is super renormalizable, and dynamical mass
generation, if it occurs at all, occurs in the deep infra red.  Two
methods of solution can be employed for the resulting approximate
integral equations: conversion to an equivalent differential equation,
or direct integration with trial functions.  The first of these will
be used with our simple approximation, and the second for the more
complete computation.

\section{Simple Computation:  Gauge Dependence.}
\label{sec:gauge}

In this section we present a computation of the wavefunction
renormalization and self energy functions based on the simplest
approximations.  In defence of the approximations, they are sufficient
to demonstrate the existence of a dynamically generated mass in
non--supersymmetric U(1) gauge theory \cite{kondo+nak92}; the
three--point vertex we choose is compatible with the Ward identity
only in the limit of vanishing transferred momentum $(p)$, though it
is consistent with a non--trivial wavefunction renormalization.  The
results of the computation for our model are clear: the expected
critical behaviour \cite{appelquist81,aitch+mav+mcneill,kondo+mura97}
is found for the wavefunction renormalization, and there is a simple
form for a gauge independent dynamically generated mass.  This is
shown in a neat way, by converting the integral equation for the mass
function to an equivalent differential equation, from which it is
obvious that a $\xi$--independent constant solution exists.  It is
impossible to use this neat method in the more complete computation to
follow in section \ref{sec:fullcomp}, for the rich structure of the
integral equation resulting from the full vertex prohibits the
conversion to a differential equation.

In the approximation $G_3 = Z(q)$ the superspace parts of the
integrals can be unpacked as described above, yielding the following
integral equations for the wavefunction renormalization and self
energy function $\left({\mathbb K}(q) \doteq Z^2(q)\,q^2 +
  \Sigma^2(q)\right)$:
\begin{eqnarray}
  \label{naivegaps}
  Z(p) &=& 1 +\frac{1}{4}\frac{\alpha}{N\pi^2} \int dq\; q^2
  \frac{Z^2(q)}{{\mathbb K}(q)} \int dx\; \frac{1}{P^4}\frac{1}{1+\alpha/|P|} 2
  (1\!+\!\xi) \left(qpx-q^2\right) , \\
\Sigma(p) &=& \frac{1}{4}\frac{\alpha}{N\pi^2} \int dq\; q^2
  \frac{\Sigma(q)Z(q)}{{\mathbb K}(q)} \int dx\;
  \frac{1}{P^4}\frac{1}{1+\alpha/|P|} \left( 2(1\!+\!\xi) \left(qpx-q^2\right) +
  2(1\!-\!\xi) P^2 \right); \nonumber
\end{eqnarray}
where $P^2 = p^2 +q^2 - 2pqx$.  The seagull graph (\ref{seagull})
gives only an irrelevant $p$ independent contribution to $Z$ which
would vanish in the supersymmetric Feynman gauge, and does not
contribute at all to $\Sigma$:
\begin{equation}
  \label{gull:trivial}
  Z_{\mathrm{seagull}} \simeq \frac{(1\!-\!\xi)}{N\pi^2} \int dq\; \left
  ( \frac{1}{q} - \frac{1}{q+\alpha} \right).
\end{equation}

The angular $x$ integrations in (\ref{naivegaps}) can be performed
easily to obtain
\begin{eqnarray}
  \label{naiveinitialgaps}
  Z(p) &=& 1 + \frac{1}{4}\frac{\alpha}{N\pi^2} \int dq\; 2q^2
  \frac{Z^2 (q) (1\!+\!\xi)}{{\mathbb K}(q)} {\mathcal I}_1
  (p,q;\alpha),\nonumber\\
  \Sigma(p) &=& \frac{1}{4}\frac{\alpha}{N\pi^2} \int dq\; 2q^2
  \frac{Z(q)\Sigma(q)}{{\mathbb K}(q)} \left[ (1\!+\!\xi){\mathcal
  I}_1(p,q;\alpha) + (1\!-\!\xi){\mathcal I}_2(p,q;\alpha) \right],
\end{eqnarray}
where the kernels ${\mathcal I}_i$ are given by
\begin{eqnarray}
  {\mathcal I}_1 &=& \frac{1}{2pq} \left( \frac{p^2-q^2}{\alpha}
  \left[ \frac{1}{|p-q|} - \frac{1}{p+q} \right] -
  \frac{p^2-q^2}{\alpha^2} \ln\left[ \frac{p+q}{|p-q|}
  \right] \right.\nonumber\\ 
  &&\qquad + \left. \frac{p^2-q^2-\alpha^2}{\alpha^2} \ln \left
  [ \frac{p+q+\alpha}{|p-q|+\alpha} \right] \right),\nonumber\\
  {\mathcal I}_2 &=& \frac{1}{pq} \ln\left
  [ \frac{p+q+\alpha}{|p-q|+\alpha} \right].
\end{eqnarray}
In the limit of $p,q \ll \alpha$ the logarithms can be expanded
\[ \ln\left[ \frac{p+q+\alpha}{|p-q|+\alpha} \right] \simeq \theta
(p-q) \left( \frac{2q}{\alpha} - 2\frac{pq}{\alpha^2} +\cdots \right)
+ \langle p\leftrightarrow q\rangle, \] and upon using the bifurcation
method in the denominators the gap equations take on the following
approximate form:
\begin{eqnarray}
  \label{naivefinalgaps}
  Z(p) &=& 1 + \frac{1}{2}\frac{(1\!+\!\xi)}{N\pi^2 \alpha p^2}
  \int_0^p dq\; q^2 - \frac{1}{2}\frac{(1\!+\!\xi)}{N\pi^2\alpha}
  \int_p^\alpha dq\; \left( 2-\frac{2\alpha}{q} - \frac{p^2}{q^2}
  \right) \nonumber\\
  \Sigma(p) &=& \frac{(1\!-\!\xi)}{N\pi^2} \int_0^p dq\; \frac{1}{p}
  \frac{\Sigma(q)}{Z(q)} - 2\frac{\xi}{N\pi^2} \int_p^\alpha dq\;
  \frac{1}{q}\frac{\Sigma(q)}{Z(q)}. 
\end{eqnarray}

\subsection{Wavefunction Renormalization.}

The first of equations (\ref{naivefinalgaps}) can be integrated
directly, to obtain the wavefunction renormalization in the normal
phase (where there is no mass generation) and as an approximation to
the wavefunction renormalization when there is mass generation:
\begin{equation}
  \label{naivewfsoln}
  Z(p) \simeq a + \left(\frac{p}{\alpha}\right)^\gamma + f\left(\frac{p}{\alpha}\right),
\end{equation}
where $a$ is a constant and the function $f$ vanishes at least as
quickly as $p/\alpha$ in the limit $p\rightarrow 0$.  In the above
we have used the usual renormalization group argument to resum
logarithmic terms as
\[ 1+ \gamma \ln \left( \frac{p}{\alpha} \right) \simeq
\left(\frac{p}{\alpha} \right)^\gamma.\] The exponent $\gamma$ is
given by
\begin{equation}
  \gamma \simeq (1\!+\!\xi) \frac{1}{N\pi^2} .
\end{equation}
This critical behaviour is qualitatively the same as that found in
non--supersymmetric electrodynamics
\cite{aitch+mav+mcneill,kondo+mura97}, so this gives us confidence in
our approach, and demonstrates that the rough qualitative behaviour is
exhibited even within the drastic approximation scheme used here.  The
non--trivial wavefunction renormalization we have found here is to be
contrasted with the early results of reference \cite{koopmans89} in
which the wavefunction renormalization was taken to be unity.

\subsection{Self Energy.}

Turning now to the self energy function, we wish to construct an
equivalent differential equation for the mass function
\[ M(p;\xi) = \frac{\Sigma(p;\xi)}{Z(p;\xi)}; \]
this is achieved by differentiating the second integral equation
(\ref{naivefinalgaps}) twice with respect to $p$ and once with respect
to $\xi$.  The reason we take an extra derivative with respect to
$\xi$ is that this will allow us to probe the gauge dependence of the
solution directly, without having to fully solve the equation, for we
are only interested in the gauge dependence of the solution.  First we
rescale variables as $p \mapsto \alpha p$ and $M(p) \mapsto \alpha
m(p)$ and then the resulting differential equation is:
\begin{equation}
  \label{naivediffeq}
  \left( \dot{m} \right)^{\prime\prime} + \frac{1}{p} \left( 2 - \frac{1}{N\pi^2}
  (1\!+\!\xi) \right) \left(\dot{m} \right)^\prime -\frac{2}{N\pi^2}
  \frac{\xi}{p^2} \dot{m} - \frac{1}{N\pi^2} \frac{1}{p} m^\prime -
  \frac{2}{N\pi^2} m = 0;
\end{equation}
the superior point denotes a derivative with respect to $\xi$ and
primes denote derivatives with respect to $p$.

Demanding that $\dot{m}=0$ for all values of $p$ (in the infra red)
yields the solution
\begin{equation}
  \label{diffsoln}
  M(p) = M_0 e^{-p^2/\alpha^2} ,
\end{equation}
which is obviously constant and $\xi$--independent in the
$p\rightarrow 0$ limit.  We should be cautious, however, for the
approximation of neglecting the mass function in the denominators
might not be consistent with this limit.  When, in the next section,
we use the full vertex and restore the mass in the denominators, we
will be able to show that this solution appears to persist in the
limit $p\rightarrow 0$, confirming our assertion here.  It is
interesting to note that there are no solutions to (\ref{naivediffeq})
which are gauge independent in the limit $p\rightarrow 0$ but gauge
dependent elsewhere (this can be shown by developing a gauge dependent
power series for $m$ and then constraining the gauge dependence via
the indicial equation): this contrasts the case of non--supersymmetric
QED${}_3$, in which such solutions do appear to exist, and seem to
lead to a critical flavour number.  We will return to a full
discussion of this issue in section \ref{sec:discuss}.

\subsection{Concluding Remarks.}

The approximations we have used in this section are rather drastic,
but the results obtained are neat and transparent:  the
wavefunction renormalization exhibits the expected critical behaviour
\cite{appelquist81,aitch+mav+mcneill,kondo+mura97}
\begin{equation}
Z(p) \simeq \left( \frac{p}{\alpha} \right)^\gamma , \qquad \gamma
\sim \frac{1}{N},
\end{equation}
and we find a gauge independent solution for the dynamically generated mass
\begin{equation}
  \label{mass:naive}
  M(p) = M_0 e^{-p^2/\alpha^2}. 
\end{equation}
These results are in line with those found in non--supersymmetric
QED${}_3$, though we have no evidence here for a critical flavour
number above which no mass generation occurs, which is the case in the
non--supersymmetric model.  This contrasts the results of reference
\cite{koopmans89}, in which a critical flavour number was found.  We
discuss this issue further in section \ref{sec:discuss} after we have
presented our full computation.

Had we not used the bifurcation method in this section, the resulting
differential equation for the mass function would have been
non--linear.  What is curious is that the bifurcation method works so
well: as we will show, the solutions derived above are entirely
consistent with the full vertex and the restoration of the mass in the
denominators of the kernel.  The inaccuracies arising from regions of
integration where the bifurcation method is least reliable are
suppressed in the limits considered for the determination of the
dynamical mass (i.e. vanishing $p$).

\section{Full Computation.}
\label{sec:fullcomp}

In this section we turn to our more complete computation, where we use
a three--point vertex consistent with the U(1) Ward identity
(\ref{wimom}).  The trick we use is to rewrite $P^2=-D^4 (P)$ in the
gauge field propagator, and then integrate by parts until the $D^2
(P)$ operators appear acting on the three--point vertex $G_3$, so that
the Ward identity (\ref{wimom}) can be used as it appears.  When this
is done, the superspace parts of the integral equations can be
unpacked as described in section \ref{sec:ds} and reconstructed after
some algebra to obtain the coupled integral equations (in which we
have already performed the angular integrations):
\begin{eqnarray}
  \label{fullgaps}
  Z(p) &=& 1 + \frac{1}{4}\frac{\alpha}{N\pi^2} \int dq\;
  \frac{q^2}{{\mathbb K}(q)} \left[ 4 (1\!+\!\xi) Z(p)Z(q) {\mathcal
  J}_1 (p,q;\alpha) \right.\nonumber\\ 
  && \qquad\qquad\qquad + \left.2(1\!-\!\xi) \left( q^2 Z(p) Z(q) -q^2 Z^2 (q) +
  \Sigma(p)\Sigma(q) - \Sigma^2(q) \right) {\mathcal J}_2 (p,q;\alpha)
  \right], \nonumber\\
  \Sigma(p) &=& \frac{1}{4}\frac{\alpha}{N\pi^2} \int dq\;
  \frac{q^2}{{\mathbb K}(q)} \left[ 2 (1\!+\!\xi) \left
  ( \Sigma(p)Z(q) - \Sigma(q) Z(p) \right) {\mathcal J}_1 (p,q;\alpha)
  \right. \nonumber\\ 
  &&\qquad\qquad\qquad + \left.2 (1\!-\!\xi) \left( q^2 \Sigma(p) Z(q)
  - p^2 \Sigma(q) Z(p) \right) {\mathcal J}_2 (p,q;\alpha) \right]; 
\end{eqnarray}
the angular integrals ${\mathcal J}_i(p,q;\alpha)$ are given by
\begin{eqnarray}
  {\mathcal J}_1 (p,q;\alpha) &=& \frac{1}{2pq} \left( \frac{p^2-q^2}{\alpha}
  \left[ \frac{1}{|p-q|} - \frac{1}{p+q} \right] -
  \frac{p^2-q^2}{\alpha^2} \ln\left[ \frac{p+q}{|p-q|}
  \right] \right. \nonumber\\ 
  && \qquad + \left.\frac{p^2-q^2-\alpha^2}{\alpha^2} \ln \left
  [ \frac{p+q+\alpha}{|p-q|+\alpha} \right] \right),\nonumber\\
  {\mathcal J}_2 (p,q;\alpha) &=& \frac{1}{pq} \left( \frac{1}{\alpha}
  \left[ \frac{1}{|p-q|} - \frac{1}{p+q} \right] - \frac{1}{\alpha^2}
  \ln\left[ \frac{p+q}{|p-q|} \right] + \frac{1}{\alpha^2} \ln\left
  [ \frac{p+q+\alpha}{|p-q|+\alpha} \right] \right).
\end{eqnarray}
Again, the seagull graph gives an irrelevant $p$ independent
contribution to $Z$, equation (\ref{gull:trivial}), and also vanishes
in the supersymmetric Feynman gauge.  Since much of our final analysis
will have to be restricted to this gauge (or at least to
approximations equivalent to considering this gauge), even if we were
to make a $p$--dependent {\sl ansatz} for $G_4$, the contribution
would vanish on account of the $C^{\alpha\beta}$ tensor structure in
the vertex projecting out only the $(1\!-\!\xi)$ component of the
gauge field propagator.

To derive the integral equations we have used the following Taylor
expansion for operators of the form $D^2(p+r)$ when acting on
$\Phi^*(p,\theta)$:
\begin{equation}
  D^2 (p+r) = D^2 (p) + \frac{1}{2} r^{\alpha\beta} \left
  [ \frac{\partial}{\partial r^{\alpha\beta}} D^2(r)\right]_{r=p} +
  \frac{1}{2} \left[ \frac{1}{2} \frac{\partial^2}{\partial
  r_{\alpha\beta} \partial r^{\alpha\beta}} D^2 (r) \right]_{r=p} . 
\end{equation}
Note that this is an exact expansion, which terminates at second
order.  The curious extra factors of one half which appear in the
expansion arise on account of the second of equations (\ref{SuSpIds})
which puts a factor of two into the dot product of two three--vectors.
When acting on $\Phi^*(p,\theta)$ we drop terms which involve
derivatives with respect to $r$, for these correspond to higher
derivative terms, which should be irrelevant for our study of the
infra red physics.

To study the integral equations it is again necessary to approximate
the logarithms with small momentum expansions, and we obtain
approximate integral equations for $\Sigma$ and $Z$, in which we have
used the abbreviated notations:
\begin{eqnarray}
  A &\doteq& \frac{1}{4} \frac{1}{N\pi^2} \nonumber\\
  Q(p,q;Z,\Sigma) &\doteq& q^2 Z(q) \left[ Z(p) - Z(q) \right] +
  \Sigma(q) \left[ \Sigma(p) - \Sigma(q) \right] \nonumber\\
  R(Z,\Sigma) &\doteq& \Sigma(p) \, Z(q) - \Sigma(q) \, Z(p)
  \nonumber\\
  S(p,q;Z,\Sigma) &\doteq& q^2 \Sigma(p) \, Z(q) - p^2 \Sigma(q), Z(p).
\end{eqnarray}
The wavefunction renormalization is given by
\begin{eqnarray}
  \label{fullgapwf}
  Z(p) = 1 + \frac{4 (1\!+\!\xi)}{\alpha p^2} &A&  \int_0^p dq\;
  \frac{q^4\, Z(p) Z(q)}{{\mathbb K}(q)}
  + 4 (1\!+\!\xi) A \int_p^\alpha dq\; \frac{Z(p) Z(q)}{{\mathbb
  K}(q)} \left( \frac{2q^2}{\alpha} - 2q - \frac{p^2}{\alpha} \right)
  \nonumber\\ 
  +2 (1\!-\!\xi) &A& \int_0^p dq\; \frac{Q(p,q;Z,\Sigma)}{{\mathbb K}(q)}
  \left( \frac{q}{p^2-q^2} -\frac{2q^2}{\alpha p^2} \right)
  \nonumber\\  
  + 2(1\!-\!\xi) &A& \int_p^\alpha dq\; \frac{Q(p,q;Z,\Sigma)}{{\mathbb
  K}(q)} \left( \frac{q^2}{p} \frac{1}{q^2-p^2} - \frac{2}{\alpha}
  \right),
\end{eqnarray}
and for the self energy we obtain
\begin{eqnarray}
  \label{fullgapse}
  \Sigma(p) = \frac{2 (1\!+\!\xi)}{\alpha p^2} &A& \int_0^p dq\;
  \frac{q^4\, R(Z,\Sigma)}{{\mathbb K}(q)}
  +2 (1\!+\!\xi) A \int_p^\alpha dq\; \frac{q^2 \,
  R(Z,\Sigma)}{{\mathbb K}(q)} \left( \frac{2}{\alpha} - \frac{2}{q} -
  \frac{p^2}{\alpha q^2} \right) \nonumber\\ 
  + 2 (1\!-\!\xi) &A& \int_0^p dq\; \frac{q^2 \, S(p,q;Z,\Sigma)}{{\mathbb
  K}(q)} \left( \frac{1}{q} \frac{1}{p^2-q^2} - \frac{2}{\alpha p^2}
  \right) \nonumber\\ 
  + 2 (1\!-\!\xi) &A& \int_p^\alpha dq\; \frac{q^2 \,
  S(p,q;Z,\Sigma)}{{\mathbb K}(q)} \left( \frac{1}{p}
  \frac{1}{q^2-p^2} - \frac{2}{\alpha q^2} \right).
\end{eqnarray}
The kernels are infra red singular; this arises from the propagation
of a supersymmetric gauge artifact \cite{clark77}, namely the lowest
component of the gauge superfield; this singular behaviour in the
infra red is a general problem for superspace formulations of gauge
field theories, but is absent in supersymmetric Feynman gauge
\cite{clark+love}.  Note that the singular terms as $p\rightarrow q$
are regulated on account of the vanishing of the functions $Q$ and $S$
in this same limit.

It is also evident from equations (\ref{fullgapwf}) and
(\ref{fullgapse}) that it is impossible to derive from them an
equivalent differential equation (not even within the bifurcation
method) on account of the presence of factors of $Z(p)$ and
$\Sigma(p)$ in the kernels.  Furthermore, it is evident that even if
the factors of $Z(p)$ and $\Sigma(p)$ could be dealt with, the same
problem arises from the $(p^2 - q^2)^{-1}$ terms: the integrals cannot
be entirely removed by differentiation.  However, this last problem is
removed in the supersymmetric Feynman gauge, in which the problem of
the gauge artifact discussed above is eliminated: for this reason much
of our analysis will be performed in Feynman gauge, or using
approximations which are computationally equivalent to this gauge.

\subsection{Wavefunction Renormalization.}
\label{sec:fullwf}

In the normal phase, where $\Sigma=0$, the integral equation for $Z$
can be integrated directly.  We begin with the approximation
$Z(p)=Z(q)$, which is not quite as drastic as $Z=1$, but has the same
effect, namely to reduce the kernels to known functions of $p$ and
$q$.  Performing the integrations yields the following form for $Z$
(compare with equation (\ref{naivewfsoln})):
\begin{eqnarray}
  Z(p) &=& 1 + \frac{2(1\!+\!\xi)}{N\pi^2} - \frac{8}{3} \frac{(1\!+\!\xi)}{N\pi^2}
  \left(\frac{p}{\alpha}\right) + \frac{(1\!+\!\xi)}{N\pi^2} \left
  ( \frac{p}{\alpha} \right)^2 + 2\frac{(1\!+\!\xi)}{N\pi^2} \ln
  \left( \frac{p}{\alpha} \right) \nonumber\\
  &\simeq& a + \left( \frac{p}{\alpha} \right)^\gamma + f\left
  ( \frac{p}{\alpha} \right) ; \label{wf:soln1}\\
  \gamma &=& 2\frac{(1\!+\!\xi)}{N\pi^2}, \label{gammadef}
\end{eqnarray}
where we have used the usual renormalization group argument to resum
the logarithm, and as in equation (\ref{naivewfsoln}), $a$ is a
constant of order ${\mathcal O}(1/N)$ and the function $f$ vanishes at
least as fast as $p/\alpha$ in the limit $p/\alpha \rightarrow 0$.
The critical exponent $\gamma$ is found to have the usual $1/N$
behaviour.  Note that the evaluation of the exponent $\gamma$ in this
more reliable approach differs by a factor of two from the result
obtained in section \ref{sec:gauge}; this is of little concern for the
exponent is not a gauge invariant object.  The $N$ dependence is the
crucial property, and this is the same in both computations.

We can demonstrate that this solution is stable, by now relaxing the
assumption that $Z(p) =Z(q)$, and feeding the solution
\[ Z(q) = \left( \frac{q}{\alpha} \right)^\gamma \] back into the integral
equation.  The parts of the integral equation which vanish in Feynman
gauge now give divergent contributions, in the form of hypergeometric
functions evaluated at the limit of their radius of convergence, and
divergent logarithms from the remaining singularities at $p=q$.  These
singularities are again a consequence of the propagation of a
supersymmetric gauge artifact; they can be successfully avoided by
considering the supersymmetric Feynman gauge, $\xi=1$, in which the
iterated solution reads:
\begin{eqnarray}
  Z(p) &=& \frac{16}{3-\gamma} \frac{1}{N\pi^2} \left( \frac{p}{\alpha}
  \right) - \frac{2}{1+ \gamma} \frac{1}{N\pi^2} \left
  ( \frac{p}{\alpha} \right)^2 + \left( \frac{p}{\alpha}
  \right)^\gamma;   \label{wf:soln2} \\
  \gamma &=& \frac{4}{N\pi^2} . 
\end{eqnarray}
The critical exponent is unaltered, and the coefficients of the other
terms receive corrections of order ${\mathcal O}(1/N)$ in the
denominator.  Hence, for large $N$ the solution is stable to
iteration, at least in supersymmetric Feynman gauge.

\subsection{Self Energy.}
\label{sec:fullse}

We turn now to the self energy, and the integral equation
(\ref{fullgapse}).  As we discussed previously, it is impossible to
derive an equivalent differential equation, on account of the presence
of $\Sigma(p)$ and \( (p^2-q^2)^{-1}\) terms in the kernels.  We again
study the equation in the Feynman gauge, which removes the
difficulties of the singular terms.  Dividing through by $Z(p)$, we
obtain the following integral equation for the mass function $M(p)$:
\begin{eqnarray}
  \left( \frac{1}{2N\pi^2} \right)^{-1} M(p) = \frac{2}{\alpha
  p^2}&& \int_0^p dq\; \frac{q^4}{Z(q)} \frac{M(p) - M(q)}{q^2 + M^2 (q)} - 4
  \int_p^\alpha dq\; \frac{q}{Z(q)}\frac{M(p)-M(q)}{q^2 +M^2(q)} \nonumber\\ 
  + \frac{4}{\alpha}&& \int_p^\alpha dq\;
  \frac{q^2}{Z(q)}\frac{M(p)-M(q)}{q^2+M^2(q)}  - \frac{2p^2}{\alpha} \int_p^\alpha
  dq\; \frac{1}{Z(q)} \frac{M(p)-M(q)}{q^2 +M^2(q)} .
\end{eqnarray}
To study this we cannot use the usual method of setting
$M(p)=M(q)=M(0)$ everywhere and looking for self consistent solutions,
for in this limit the kernels reduce to zero.  Instead, we recall the
solution of the differential equation in section \ref{sec:gauge},
equation (\ref{diffsoln}), and adopt the following {\sl ansatz} for
$M$ inside the kernels:
\begin{equation}
  \label{Mansatz}
  M(q) = m_0 e^{-p^2/\alpha^2}.
\end{equation}
We adopt also the solution \[Z(q) = \left( \frac{q}{\alpha}
\right)^\gamma \] for the wavefunction renormalization factors which
appear above.  Unfortunately the integrations cannot be performed
analytically if the exponential factor in (\ref{Mansatz}) is present
in the denominators.  Hence in the denominators we set $M$ to a
constant, $\bar{m}$, and it is with this in mind that the scale
$\bar{m}$ will be interpreted as an average mass over the range $p\in
[0,\alpha]$; we will see that in the limits we wish to consider,
dependence on the introduced scale drops out.  We can then perform the
integrations analytically with result:
\begin{eqnarray}
  \label{fullresult}
  \frac{M(p)}{m_0} &=& -\frac{\gamma \pi}{4} \left( \frac{\alpha}{\bar{m}}
  \right)^\gamma e^{-p^2/\alpha^2} \csc \left( \frac{2-\gamma}{2} \pi
  \right) + \frac{1}{2} e^{-p^2/\alpha^2} - \frac{\gamma}{4} \Gamma \left
  ( -\frac{\gamma}{2} , \left(\frac{\bar{m}}{\alpha}\right)^{3/2}
  \right) \nonumber\\
&& + \frac{\gamma}{4} \left( \frac{\alpha}{\bar{m}}
  \right)^\gamma e^{\bar{m}^2/\alpha^2} \Gamma \left
  ( \frac{2-\gamma}{2} \right) \Gamma \left( \frac{\gamma}{2} ,
  \frac{\bar{m}^2}{\alpha^2} \right) 
- \frac{\gamma}{2} \frac{1}{1-\gamma} \left( \frac{p}{\alpha}
  \right)^{1-\gamma} e^{-p^2/\alpha^2} \nonumber\\
&& - \frac{1}{2} \frac{\gamma}{2-\gamma} \left
  ( \frac{\alpha}{\bar{m}} \right)^{\gamma} \left
  ( \sqrt{\frac{p}{\bar{m}}}\right)^{(2-\gamma)/2} {}_2F_1 \left( 1,
  \frac{2-\gamma}{2} ; \frac{4-\gamma}{2}, - \sqrt{\frac{p}{\bar{m}}}
  \right)\nonumber\\
&& + \frac{\gamma}{2} \frac{1}{1-\gamma} e^{-p^2/\alpha^2}
- \frac{\gamma\pi}{4} \left( \frac{\bar{m}}{\alpha} \right)^{1-\gamma}
  e^{-p^2/\alpha^2} \csc \left( \frac{1-\gamma}{2} \pi \right)
+ \frac{1}{2} \frac{\gamma}{1-\gamma} \left( \frac{p}{\alpha}
  \right)^{1-\gamma} e^{-p^2/\alpha^2}\nonumber\\ 
&& - \frac{1}{2} \frac{\gamma}{1+\gamma} \left( \frac{\bar{m}}{\alpha}
  \right)^2 e^{-p^2/\alpha^2}
- \frac{\gamma}{4} \left( \frac{\bar{m}}{\alpha}
  \right)^{1-\gamma} e^{\bar{m}^2/\alpha^2} \Gamma\left
  ( \frac{3-\gamma}{2} \right) \Gamma \left( \frac{\gamma-1}{2} ,
  \frac{\bar{m}^2}{\alpha^2} \right) \nonumber\\
&& + \frac{1}{2} \frac{\gamma}{3-\gamma} \left( \frac{\bar{m}}{\alpha}
  \right)^{1-\gamma} \left( \sqrt{\frac{p}{\bar{m}}}
  \right)^{(3-\gamma)/2} {}_2F_1 \left( 1, \frac{3-\gamma}{2} ;
  \frac{5-\gamma}{2} , - \sqrt{\frac{p}{\bar{m}}} \right) \nonumber\\
&& + \frac{\gamma}{4} \Gamma \left( \frac{1-\gamma}{2} , \left
  ( \frac{\bar{m}}{\alpha} \right)^{3/2} \right)
+ {\mathcal O} \left ( \frac{p^2}{\alpha^2} \right),
\end{eqnarray}
where \[ \gamma = \frac{4}{N\pi^2}.\]  The limit $p\rightarrow 0$ is
now well defined, and if we also assume \( \bar{m} \ll \alpha \) and
take the limit \( \gamma \ll 1 \) we find that
\begin{equation}
  \label{fullsoln}
  \frac{M(0)}{m_0} = \frac{1}{1-\gamma} + \frac{\gamma}{2} \left
  ( \frac{\pi}{2} - \frac{1}{1-\gamma} \right) + \ldots
\end{equation}
where the dots represent the omission of terms smaller than the finite
terms exhibited above.  It is clear then, that to leading order in
$1/N$, the {\sl ansatz} (\ref{Mansatz}) is stable to iteration, and we
have found that the solution of the differential equation in section
\ref{sec:gauge} is consistent with the full vertex and the restoration
of the mass in the denominators of the kernels.  We have therefore
demonstrated the possibility of a finite dynamically generated mass
for this theory, though we still find no evidence for a critical
flavour number: we return to a discussion of this issue in section
\ref{sec:discuss}.

\subsection{Concluding Remarks.}

In this section we have presented a computation of the wavefunction
renormalization and self energy with a three-point vertex which is
consistent with the U(1) Ward identity.  While we have not been able
to use the elegant methods of section \ref{sec:gauge}, we have been
able to perform a direct computation, and within the approximations
made we have found that the wavefunction renormalization exhibits the
expected critical behaviour
\cite{appelquist81,aitch+mav+mcneill,kondo+mura97} and we have
exhibited a finite solution to the gap equation for the mass function:
\begin{eqnarray}
  \label{fullconcs}
  Z(p) &\simeq& \left( \frac{p}{\alpha} \right)^\gamma, \qquad \gamma =
  2\frac{(1\!+\!\xi)}{N\pi^2}; \nonumber\\ &&\nonumber\\  
  M(p) &=& m_0 e^{-p^2/\alpha^2}.
\end{eqnarray}
These results are qualitatively the same as those derived using the
simple approximations of section \ref{sec:gauge}, justifying the
use of those approximations in determining qualitatively the gauge
dependence in this model.

Since we have found that a mass can be dynamically generated in this
model without breaking supersymmetry, it is natural to ask whether it
is possible to determine whether the massive or massless (for $m_0$
could still vanish) solution is favoured.  In non--supersymmetric
theories this question can be addressed by appealing to the effective
potential (which upon using a suitable variational principle leads to
the same non--perturbative physics for QED${}_3$ as do the
Dyson--Schwinger equations \cite{adrian98}), but in supersymmetric
models this potential vanishes.  Based on experience with
supersymmetric non--linear sigma models \cite{alvarez78} it was
conjectured \cite{pisarski84} that the question could be settled by
appealing to the effective action, and to the Ward identities arising
from the supersymmetry.  The same approach was taken in reference
\cite{diamandis98} for a ($2+1$ dimensional) supersymmetric model with
a CP${}^1$ constraint.  The simplest Ward identity arising from
supersymmetry relates the two--point correlation functions of the
physical degrees of freedom in the matter superfield (in the presence
of a mass $m$ for the multiplet):
\begin{equation}
  \label{comptwi}
  \langle \psi_\alpha (x) \psi_\beta (y) \rangle_0 = \left
  ( \partial_{\alpha\beta} + m \,C_{\alpha\beta} \right) \langle
  \phi(x) \phi(y) \rangle_0 .
\end{equation}
Crucially, in the models of references
\cite{alvarez78,ciuchini95,diamandis98}, where there are further
similar Ward identities relating correlation functions of fields
arising from the implementation of the constraints, it was found that
these Ward identities were only satisfied when the mass for the matter
multiplet was non--vanishing.  In the present model, and in the
${\mathcal N}=2$ model of reference \cite{pisarski84}, there are no
such constraints, and so the Ward identities place no further
requirements on the generated mass; of course, the superfield
formalism we have adopted means that the Ward identity (\ref{comptwi})
is satisfied automatically.  This leaves open the question of whether
the vacuum selects the massless or massive solution in supersymmetric
U(1) gauge field theory.

\section{Extended Supersymmetric Action.}
\label{sec:n=2action}

The ${\mathcal N}=2$ extended rigid superspace is constructed in a way
slightly different to the ${\mathcal N}=1$ model: it can be obtained
by dimensional reduction from ${\mathcal N}=1$ supersymmetry in four
dimensions.  The compactification is realised by considering all the
fields in the model to be independent of one spatial dimension and
then integrating out this variable.  The rigid superspace of four
dimensional ${\mathcal N}=1$ supersymmetry is parameterized by the
usual four space--time coordinates and four Grassmann--odd coordinates
arranged in two spinors $\theta^\alpha$ and
$\bar{\theta}^{\dot{\alpha}}$, where $\alpha \in \{ 1,2 \}$ and
$\dot{\alpha} \in \{ \dot{1},\dot{2} \}$ \cite{srivastava}, which we
again denote collectively by $z$.  This furnishes the superspace with
two spinorial derivatives $D_\alpha$ and $\bar{D}_{\dot{\alpha}}$,
covariant under supersymmetry transformations, which leads to a
crucial difference compared to the three dimensional model, in that
the matter superfields now come in two types: chiral $(\Phi)$ and
antichiral $(\Phi^\dagger)$, which respectively obey the constraints
\begin{equation}
  \label{chiral}
  \bar{D}_{\dot{\alpha}} \Phi = 0 , \qquad D_\alpha \Phi^\dagger = 0.
\end{equation}
We collect some basic results for four dimensional superspace in
appendix \ref{app:4d}.  Consider chiral and antichiral superfields
which transform under local U(1) transformations in the following way:
\begin{eqnarray}
  \Phi_\pm (z) \longrightarrow &\Phi^\prime_\pm (z^\prime)& = e^{\pm
  ie \Lambda(z)} \Phi_\pm (z),\nonumber\\
  \Phi^\dagger_\pm (z) \longrightarrow &\Phi^{\dagger\prime}_\pm
  (z^\prime)& =  \Phi^\dagger_\pm (z) e^{\mp ie \bar{\Lambda(z)}}.
\end{eqnarray}
The functions $\Lambda$ and $\bar{\Lambda}$ must respectively be
chiral and antichiral, to preserve the chiral and antichiral nature of
the matter superfields.  Note that $e$ is a dimensionless gauge
coupling: the dimensionful parameter which will again lead to a
dynamically generated scale will be the size of the compactified
dimension.  The gauge field is now a real scalar superfield, and
transforms under infinitesimal U(1) transformations in the following
way
\begin{equation}
  V(z) = V^\dagger (z) \longrightarrow V^\prime (z^\prime) = V (z) +
  \frac{i}{2} \left( \bar{\Lambda} - \Lambda \right).
\end{equation}
We now wish to consider a model with ${\mathcal N}=1$ supersymmetry,
local U(1) gauge invariance and $N$ matter flavours; again the action
functional comprises three parts, a locally U(1) invariant kinetic
term for the matter superfields, a gauge invariant classical field
strength for the gauge superfield, a (Lorentz) gauge fixing term; for
convenience we will also add a bare mass term for the matter
superfields:
\begin{equation}
  \label{n=2action}
  S = S_{g}^{\mathrm{class}} + S_{g}^{\mathrm{GF}} +
  S_m^{\mathrm k} + S_m^{\mathrm m};
\end{equation}
\begin{eqnarray}
  S_{g}^{\mathrm{class}} &=& \int d^4 x\,d^4\theta\; V
  \left[ \frac{1}{8} D^\alpha \bar{D}^2 D_\alpha \right] V,\nonumber\\
  S_g^{\mathrm{GF}} &=& \int d^4 x\,d^4\theta\; \frac{1}{8} V \left
  [ -\frac{1}{8\xi} D^2 \bar{D}^2 \right] V,\nonumber\\
  S_m^{\mathrm k} &=& \int d^4 x\,d^4\theta\; \left[ \Phi^\dagger_+ e^{2eV}
  \Phi_+ + \Phi^\dagger_- e^{-2eV} \Phi_- \right], \nonumber\\
  S_m^{\mathrm m} &=& \int d^4 x\, d^2 \theta\; \mu_0 \,\Phi_+ \Phi_- + \int
  d^4x \, d^2\bar{\theta}\; \mu_0 \,\Phi^\dagger_+ \Phi^\dagger_- .
\end{eqnarray}
Again we have included an implicit sum over $N$ flavours in the matter
parts.  There are a number of important differences between this
action and the one we considered in the first part of this paper.
First, there are now twice as many matter superfields: an antichiral
and chiral superfield for each charge under U(1): as we shall see this
changes considerably the structure of the propagators; in particular,
note that in the absence of a bare mass $\mu_0$ the non--interacting
correlation function \( \langle \Phi_+ \Phi_- \rangle_0 \) vanishes,
and the Dyson--Schwinger equations for the self energy and
wavefunction renormalization are decoupled to an extent.  Furthermore,
it is evident that the implementation of the interaction with the
gauge field is completely different: there are now an infinite number
of vertices with increasing numbers of gauge superfields.  In the
Wess--Zumino gauge (which projects out the physical components of the
gauge superfield, but which thereby breaks supersymmetry explicitly)
the sequence of interaction vertices terminates at the four point
vertex, giving it the same content as the model considered in the
first part of this paper.  In order to make the model tractable, we
will truncate the Dyson--Schwinger equations at the level of the
four--point vertex, effectively setting higher--point vertices to zero
\cite{clark+love}.  This situation is certainly not ideal
\cite{kaiselip}, but we find that the graph in the Dyson--Schwinger
equation arising from the four--point vertex contributes trivially to
the wavefunction renormalization, and we therefore believe that this
truncation will not affect the infra red physics.  The dressed
propagators for the model are as follows:
\begin{eqnarray}
  \label{n=2props}
  \Delta^{\Phi_+ \Phi_-} (p;12) &=& \frac{i\Sigma(p)}{Z^2(p)\, p^2 +
\Sigma^2(p)}\frac{\bar{D}^2}{4} \delta^4(12),\nonumber\\
\Delta^{\Phi_{\pm}^{\dagger} \Phi_{\pm}}(p;12) &=& \frac{i
Z(p)}{Z^2(p)\, p^2 + \Sigma^2(p)}\frac{D^2 \bar{D}^2}{16}
\delta^4(12),
\end{eqnarray}
we have again used an abbreviated superspace notation where, for
example \[ \delta^4(12)\equiv\delta^4(\theta_1 - \theta_2) =
\delta^2(\theta_1 -\theta_2) \, \delta^2(\bar{\theta}_1
-\bar{\theta}_2) \equiv \delta^2(12) \, \delta^2 (\bar{1}\bar{2}).\]
The undressed gauge superfield propagator reads \cite{srivastava}
\begin{equation}
  \Delta^{VV}(p;12)= \frac{1}{2p^2}e^{\theta_1\slp\bar{\theta}_2 -
\theta_2\slp\theta_1} \left\{\frac{4}{p^2}\left( 1-\xi \right) -
\left(1+\xi\right) \delta^4(12) \right\};
\end{equation}
later we will dress this propagator with a vacuum polarization factor
as before.

For simplicity we will restrict our attention to the simple vertex
approximation considered in the approach of section \ref{sec:gauge};
this is in fact consistent with the U(1) Ward identity for this model
\cite{clark+love}, at least in the limit of vanishing transferred
momentum $p$.

\section{The Dyson--Schwinger Equations.}
\label{sec:n=2ds}

We can rewrite the matter superfields in terms of unconstrained scalar
superfields, which makes the evaluation of the superspace parts of the
Dyson--Schwinger equations simpler:
\begin{eqnarray}
  \Phi_\pm (p,\theta,\bar{\theta}) &=& e^{-\theta \slp \bar{\theta}}
\Psi_\pm (p,\theta),\nonumber\\
  \Phi_\pm^{\dagger} (p,\theta,\bar{\theta}) &=& e^{\theta \slp \bar{\theta}}
\Psi_\pm^{\dagger} (p,\bar{\theta}).
\end{eqnarray}

\begin{figure}
\begin{center}
\begin{picture}(280,80)(0,0)
\Line(0,30)(60,30)
\GCirc(30,30){7}{0.5}
\LongArrow(20,20)(40,20)
\Text(30,10)[]{$p$}
\Text(5,40)[]{$\Phi_+$}
\Text(55,40)[]{$\Phi_+^{\dagger}$}
\Text(70,30)[]{$=$}
\Text(85,30)[]{$-$}
\Line(100,30)(160,30)
\PhotonArc(130,48)(15,0,360){3.5}{7}
\LongArrow(120,20)(140,20)
\Vertex(130,30){4}
\LongArrowArcn(130,44)(12,130,50)
\Text(130,48)[]{$q$}
\Text(130,10)[]{$p$}
\Text(105,40)[]{$\Phi_+$}
\Text(155,40)[]{$\Phi_+^{\dagger}$}
\Text(170,30)[]{$-$}
\Line(180,30)(280,30)
\GCirc(230,30){7}{0.5}
\PhotonArc(230,30)(25,5,180){4}{5.5}
\LongArrowArcn(230,38)(25,130,50)
\Text(232,72)[]{$p-q$}
\Vertex(205,30){1}
\Vertex(255,30){4}
\Text(185,40)[]{$\Phi_+$}
\Text(275,40)[]{$\Phi_+^{\dagger}$}
\LongArrow(220,20)(240,20)
\Text(230,10)[]{$q$}
\end{picture}
\caption{\label{fig:n=2wf}Schematic form of the Dyson--Schwinger
  equation for the wave-function renormalization.  Solid lines
  represent matter superfields and wavy lines represent gauge
  superfields; blobs indicate full non--perturbative quantities.}
\end{center}
\end{figure}
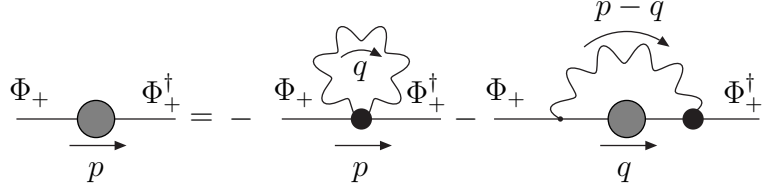

The Dyson--Schwinger equations for the current model are shown
schematically in figures \ref{fig:n=2se} and \ref{fig:n=2wf}.  The
graphs on the left hand side are a convenient abbreviation for the
difference between the inverse full propagator and inverse bare
propagator.

\begin{figure}
\begin{center}
\begin{picture}(200,80)(0,0)
\Line(0,30)(60,30)
\GCirc(30,30){7}{0.5}
\LongArrow(20,20)(40,20)
\Text(30,10)[]{$p$}
\Text(5,40)[]{$\Phi_+$}
\Text(55,40)[]{$\Phi_-$}
\Text(70,30)[]{$=$}
\Text(85,30)[]{$-$}
\Line(100,30)(200,30)
\GCirc(150,30){7}{0.5}
\PhotonArc(150,30)(25,5,180){4}{5.5}
\LongArrowArcn(150,38)(25,130,50)
\Text(152,72)[]{$p-q$}
\Vertex(125,30){1}
\Vertex(175,30){4}
\Text(105,40)[]{$\Phi_+$}
\Text(195,40)[]{$\Phi_-$}
\LongArrow(140,20)(160,20)
\Text(150,10)[]{$q$}
\end{picture}
\caption{\label{fig:n=2se}Schematic form of the Dyson--Schwinger
  equation for the self energy function.  Solid lines represent matter
  superfields, and the wavy line represents the gauge superfield;
  blobs indicate full non--perturbative quantities.}
\end{center}
\end{figure}
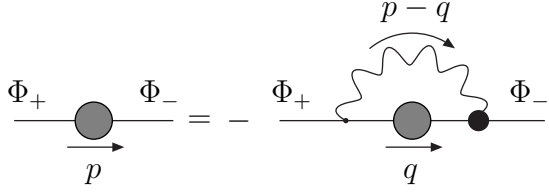

The full self energy function on the left hand side of figure
\ref{fig:n=2se} is written as follows:
\begin{eqnarray}
  \int&& d^2 \theta_1 \, d^2 \theta_2 \; \Phi_+ (-p,\theta_2) \,\Sigma(p)
\frac{\bar{D}^2}{4} \delta^4(12) \, \Phi_-(p,\theta_1) \nonumber\\
&&= \int d^2\theta_1 \, d^2 \theta_2 \; e^{\theta_2\slp\bar{\theta}_2}
\Psi_+ (-p,\theta_2) \, \Sigma(p) \delta^2(12) \, e^{-\theta_1\slp
(\bar{\theta}_1 - \bar{\theta}_2)}
e^{-\theta_1\slp\bar{\theta}_1}\Psi_-(p,\theta_1)\nonumber\\
&&= \int d^2 \theta \; \Psi_+(-p,\theta) \Sigma(p) \Psi_- (p, \theta).
\end{eqnarray}
The graph on the right of figure \ref{fig:n=2se} is given by the
following $( P \doteq p-q)$:
\begin{eqnarray}
\int&& \frac{d^4q}{(2\pi)^4} \, d^4\theta \, d^4\bar{\theta} \; \Phi_+
(-p,\theta_2) \,\left[ \left( -4e^2 G_3\right) \,\Delta^{\Phi_+^{\dagger}
\Phi_-^{\dagger}} (q;12) \,\Delta^{VV}(P;12) \right],\Phi_- (p,\theta_1)
\nonumber\\
&&= -\int \frac{d^4q}{(2\pi)^4} \, d^4\theta \, d^4\bar{\theta}
\; 4e^2 G_3 \, \Phi_+ (-p,\theta_2) \, \frac{i\Sigma(q)}{Z^2(q)\, q^2+\Sigma^2(q)}
\frac{D^2}{4}\delta^4(12) \nonumber\\
&& \quad \times \frac{1}{2P^2}\, e^{\theta_1 \not{P}
\bar{\theta}_2 -\theta_2 \not{P}\bar{\theta}_1}
\left\{\frac{4}{P ^2}(1\!-\!\xi) - (1\!+\!\xi)\delta^4(12)\right\}
\Phi_-(p,\theta_1).
\end{eqnarray}
Expanding the exponentials and performing half of the superspace
integrations using the identities in appendix \ref{app:4d} leaves:
\begin{equation}
i \int \frac{d^4q}{(2\pi)^4} \; 4e^2 \, \Gamma_3 \,
\frac{\Sigma(q)}{Z^2(q)\, q^2+\Sigma^2 (q)} \,
\frac{2p^2}{P^4}(1\!-\!\xi)\int d^2\theta \; \Psi_+(-p,\theta)\Psi_-(p,\theta).
\end{equation}
Note that in this model the superspace integrations project out the
$(1\!-\!\xi)$ component of this graph.  This is the basis of the
gauge dependence argument \cite{clark+love} for the non--generation of
dynamical mass in this theory.

Turning to figure \ref{fig:n=2wf}, the graph on the left for the full
wavefunction renormalization is written as follows:
\begin{eqnarray}
\int&& d^2 \theta_1\, d^2 \bar{\theta}_2 \; \Phi_+(-p,\theta_2) \,
\left( Z(p) -1 \right) \, \frac{\bar{D}^2 D^2}{16}\, \delta^4(12)\,
\Phi_+^{\dagger}(p,\theta_1) \nonumber \\
&&=\int d^2 \theta_1\, d^2 \bar{\theta}_2 \;
e^{\theta_2\slp\bar{\theta}_2} \Psi_+(-p,\theta_2) \,
\left( Z(p) -1\right) \, e^{-\theta_1\slp\bar{\theta}_1
-\theta_2\slp\bar{\theta}_2 +2\theta_1\slp\bar{\theta}_2} \,
e^{\theta_1\slp\bar{\theta}_1} \, \Psi_+^{\dagger}(p,\bar{\theta}_1)\nonumber\\
&&=\int d^2 \theta_1\, d^2 \bar{\theta}_2 \; \Psi_+
(-p,\theta_2) \left( Z(p) -1 \right) \,
e^{2\theta_1\slp\bar{\theta}_2} \,
\Psi_+^{\dagger}(p,\bar{\theta}_1)\nonumber\\
&&=\int d^4 \theta\; (Z(p)-1) \, \Phi_+(-p,\theta,
\bar{\theta})\, \Phi_+^{\dagger}(p,\theta,\bar{\theta}).
\end{eqnarray}
The superspace integrations can be performed in the same way as for
the self energy to write the graphs on the right of figure
\ref{fig:n=2wf}.  The ``seagull'' graph is computed to be a trivial
$p$ independent contribution, as in the model of the first part of
this paper (and which again vanishes in supersymmetric Feynman gauge):
\begin{eqnarray}
  \int&& \frac{d^4q}{(2\pi)^4} \; d^4\theta \; \frac{1}{2} 4e^2 G_4
\, \Phi_+(-p,\theta,\bar{\theta})\, \Delta^{VV}(q)\,
\Phi_+^{\dagger}(p,\theta,\bar{\theta})\nonumber\\
&&=\int \frac{d^4q}{(2\pi)^4} \; 2e^2\Gamma_4\,
\frac{2}{P^4}(1\!-\!\xi) \int d^4\theta \;
\Phi_+^{\dagger}(p,\theta,\bar{\theta})\, \Phi_+(-p,\theta,\bar{\theta}).
\end{eqnarray}
The last graph on the right of figure \ref{fig:n=2wf} is as follows:
\begin{eqnarray}
  \int&& \frac{d^4q}{(2\pi)^4} \; d^4\theta_1 \, d^4 \theta_2 \;
\Phi_+(-p,\theta_2) \, 4e^2 G_3 \,
\Delta^{\Phi_+^{\dagger}\Phi_+}(q;12) \, \Delta^{VV}(P;12) \,
\Phi_+^{\dagger}(p,\theta_1) \nonumber \\
&&=\int \frac{d^4q}{(2\pi)^4} \; 4e^2 G_3 \, \frac{iZ(q)}{Z^2(q)\, q^2
+ \Sigma^2(q)}\, \left[ \frac{(p+q)^2}{2 P^4}(1\!-\!\xi) - (1\!+\!\xi)
\right]\nonumber \\
&&\quad \times \int d^4\theta \; \Phi_+(-p,\theta,\bar{\theta}) \,
\Phi_+^{\dagger}(p,\theta_1).
\end{eqnarray}

Collecting the results above, the Dyson--Schwinger equations of
figures \ref{fig:n=2se} and \ref{fig:n=2wf} then yield the following
coupled integral equations for the self energy and wavefunction
renormalization:
\begin{eqnarray}
  \label{n=2segap1}
  \Sigma(p) = &&-i\, 4e^2 p^2 (1\!-\!\xi) \int \frac{d^4q}{(2\pi)^4} \;
G_3 \frac{\Sigma(q)}{Z^2(q)\, q^2 + \Sigma^2(q)}\frac{2}{(p-q)^4};\\
Z(p) =  &1& -i \, 4e^2 \, (1\!+\!\xi)\int \frac{d^4q}{(2\pi)^4} \; G_3
\frac{Z(q)}{Z^2(q)\, q^2 + \Sigma^2(q)}\frac{1}{2(p-q)^2} \nonumber \\
&&+i4e^2 \, (1\!-\!\xi)\int \frac{d^4q}{(2\pi)^4} \; G_3
\frac{Z(q)}{Z^2(q)\, q^2 + \Sigma^2(q)} \frac{(p+q)^2}{(p-q)^4} \nonumber \\
&&+ 4e^2 (1\!-\!\xi) \int \frac{d^4q}{(2\pi)^4} \; G_4
\frac{1}{q^4}. \label{wfgap1}
\end{eqnarray}

To study the infra red physics of ${\mathcal N}=2$ supersymmetric U(1)
gauge field theory in three dimensions, we will simply consider all
the fields above to be independent of one coordinate, and then
integrate it out: this will introduce a natural scale $\rho$, the size
of the compactified dimension, and we will then be able to compare
directly with the model of the first part of this paper.  We
compactify one dimension as follows:
\begin{equation}
\int dq_3 = \frac{1}{\rho}
\end{equation}
We can now incorporate the effects of massless matter loops into the
gauge field propagator, by dressing it with the appropriate vacuum
polarization factor:
\begin{equation}
  \Delta^{VV}(p;12) \longrightarrow \frac{\Delta^{VV}(p;12)}{1 + {\bar{\alpha}}/|p|},
\end{equation}
where we have introduced the ultra violet scale \[ {\bar{\alpha}}
\doteq \frac{e^2 N}{\rho}\] in analogy with the treatment of normal
QED${}_3$ and the ${\mathcal N}=1$ model considered in the first part
of this paper.  The combination $e^2/\rho$ relates the dimensionless
coupling $e$ of the original model to the dimensionful coupling in the
dimensionally reduced theory.  We can carry out the angular part of
the momentum integration easily in three dimensions, with result
\begin{eqnarray}
\Sigma (p) =&& \frac{{\bar{\alpha}}}{2N\pi^3} (1\!-\!\xi) \int dq \; G_3
\frac{\Sigma(q)}{Z^2(q) \, q^2 + \Sigma^2 (q)} \, 2qp \, \left(
\frac{1}{{\bar{\alpha}}} \left( \frac{1}{|p-q|}-\frac{1}{p+q}\right)
\right. \nonumber\\
&& \qquad\qquad\qquad \left.+ \frac{1}{{\bar{\alpha}}^2} \ln \left[
\frac{p+q+{\bar{\alpha}}}{|p-q|+{\bar{\alpha}}}\right] -\frac{1}{{\bar{\alpha}}^2}\ln \left[
\frac{p+q}{|p-q|}\right] \right);\nonumber \\
Z(p) = &1& + \frac{{\bar{\alpha}}}{2N\pi^3} (1\!+\!\xi) \int dq \; G_3
\frac{Z(q)}{Z^2(q) \, q^2 + \Sigma^2 (q)} \,\frac{q}{2p}\, \ln
\left[\frac{p+q+{\bar{\alpha}}}{|p-q|+{\bar{\alpha}}}\right] \nonumber \\
&& - \frac{{\bar{\alpha}}}{2N\pi^3} (1\!-\!\xi) \int dq \; G_3
\frac{Z(q)}{Z^2(q) \, q^2 + \Sigma^2 (q)} \,\frac{q}{2p}\, \left(
\frac{2(p^2+q^2)}{{\bar{\alpha}}} \left( \frac{1}{|p-q|} -
\frac{1}{p+q}\right) \right. \nonumber\\
&& \qquad\qquad\qquad \left.-\frac{2(p^2+q^2)}{{\bar{\alpha}}^2} \ln \left[
\frac{p+q}{|p-q|}\right] +
\frac{2(p^2+q^2)-{\bar{\alpha}}^2}{{\bar{\alpha}}^2} \ln \left[
\frac{p+q+{\bar{\alpha}}}{|p-q|+{\bar{\alpha}}}\right] \right) \nonumber\\
&& -\frac{i{\bar{\alpha}}}{2N\pi^3} (1\!-\!\xi) \int dq \; \frac{q}{p}
G_4 \left( \frac{1}{{\bar{\alpha}}} \left(\frac{1}{|p-q|} -
\frac{1}{p+q}\right) \right. \nonumber\\
&& \qquad\qquad\qquad \left. -\frac{1}{{\bar{\alpha}}^2} \ln \left[
\frac{p+q}{|p-q|}\right] +\frac{1}{{\bar{\alpha}}^2} \ln \left[
\frac{p+q+{\bar{\alpha}}}{|p-q|+{\bar{\alpha}}}\right] \right).
\end{eqnarray}
To study these equations it is once again necessary to expand the
logarithms arising from the angular integrations, in the limit $p,q
\ll {\bar{\alpha}}$, which leaves
\begin{eqnarray} \label{n=2gaps}
  \Sigma(p) =&& \frac{{\bar{\alpha}}}{N\pi^3} (1\!-\!\xi) \int dq \; {\mathcal{F}}(q)
\left[ \frac{2q^2 p}{p^2-q^2} \theta(p-q) + \frac{2 p^2 q}{q^2-p^2}
\theta(q-p) \right],\nonumber\\ 
Z(p) = &1& + \frac{{\bar{\alpha}}}{2N\pi^3} (1\!+\!\xi) \int dq\;
{\mathcal{G}}(q) \left[  \frac{q^2}{{\bar{\alpha}} p}
\theta(p-q) + \frac{q}{{\bar{\alpha}}} \theta(q-p) \right]\nonumber\\
&& - \frac{{\bar{\alpha}}}{2N\pi^3} (1\!-\!\xi) \int dq\; {\mathcal{G}}(q)
\left[ \left( \frac{p^2+3q^2}{p^2-q^2}\right)\frac{q^2}{{\bar{\alpha}} p}
\theta(p-q) + \left( \frac{3p^2+q^2}{q^2-p^2}\right) \frac{q}{{\bar{\alpha}}}
\theta(q-p)\right] \nonumber\\ 
&& - \frac{i{\bar{\alpha}}}{2N\pi^3} (1\!-\!\xi) \int dq\; \Gamma_4 \left[
\frac{2q^2}{{\bar{\alpha}} p}\frac{1}{p^2-q^2} \theta(p-q) +
\frac{2q}{{\bar{\alpha}}}\frac{1}{q^2-p^2} \theta(q-p) \right],
\end{eqnarray}
where we have introduced the convenient abbreviated notations:
\begin{eqnarray}
  {\mathcal{F}}(q) = G_3 \, \frac{\Sigma(q)}{Z^2 (q) \, q^2 +
\Sigma^2 (q)} \nonumber\\
{\mathcal{G}}(q)= G_3 \, \frac{Z(q)}{Z^2(q) \, q^2 +\Sigma^2(q)}.
\end{eqnarray}

\subsection{Wavefunction Renormalization.}
\label{sec:n=2wf}

In the normal phase (or equivalently, within the bifurcation method)
where $\Sigma=0$, the integral equation for the wavefunction
renormalization can be integrated directly upon using the simple
approximation $G_3 = Z(q)$, $G_4=1$ \cite{clark+love}:
\begin{eqnarray}
  \label{n=2wfnsoln}
  Z(p)= &1&+ \frac{{\bar{\alpha}}}{2N\pi^3} (1\!+\!\xi) \left[ \frac{1}{{\bar{\alpha}}} -
\frac{1}{{\bar{\alpha}}}\ln \left( \frac{p}{{\bar{\alpha}}}\right) \right]\nonumber \\
&& - \frac{{\bar{\alpha}}}{2N\pi^3} (1\!-\!\xi) \left[ -\frac{3}{{\bar{\alpha}}} +
\frac{4}{{\bar{\alpha}}}{\tanh}^{-1} \left( 1-\frac{\varepsilon}{p}\right)
\right]\nonumber\\
&& - \frac{{\bar{\alpha}}}{2N\pi^3} (1\!-\!\xi) \left[ \frac{3}{{\bar{\alpha}}} \ln
\left(\frac{p}{{\bar{\alpha}}}\right) + \frac{2}{{\bar{\alpha}}}\ln\left(
\frac{{\bar{\alpha}}^2 - p^2}{2p\varepsilon}\right)\right]\nonumber\\
&& - \frac{i{\bar{\alpha}}}{2N\pi^3} (1\!-\!\xi) \left[  -\frac{2}{{\bar{\alpha}}} +
\frac{2}{{\bar{\alpha}}}\tanh^{-1}\left(1-\frac{\varepsilon}{p}\right)
+\frac{1}{{\bar{\alpha}}}\ln\left( \frac{{\bar{\alpha}}^2 -p^2}{2p\varepsilon}\right)  \right] ,
\end{eqnarray}
where the limit $\varepsilon\rightarrow 0^+$ is implied.  Again,
outside the supersymmetric Feynman gauge, there are many infra red
divergences which result from the propagation of a gauge artifact in
the gauge superfield.  However, ignoring these divergences as
spurious, we can see that the solution above is consistent with the
expected critical behaviour, with an exponent of order ${\mathcal O}(1/N)$:
\begin{equation}
  \label{n=2critwf}
  Z(p) \simeq \left( \frac{p}{{\bar{\alpha}}}
  \right)^{\bar{\gamma}} \qquad \bar{\gamma} \simeq \frac{1-2\xi}{N\pi^3}.
\end{equation}
While it is not obvious that this form for the wavefunction
renormalization is completely reliable, what is clear is that there is
no evidence for an overall factor of $(1\!-\!\xi)$ in the
wavefunction renormalization, a fact which will be crucial for the
discussion of the self energy which follows.

\subsection{Self Energy.}
\label{sec:n=2se}

It was argued in reference \cite{clark+love} that the overall factor
of $(1\!-\!\xi)$ in the first of equations (\ref{n=2gaps}) and its
four dimensional ancestor (\ref{n=2segap1}) for the self energy means
that the self energy vanishes in Feynman gauge, and since the full
gauge dependence was carried through, this indicates that the self
energy vanishes in all gauges; this was conjectured to be the result
of a four dimensional (non--perturbative) non--renormalization theorem
\cite{clark+love}.  This argument has been criticized
\cite{kaiselip,appelquist98} on the grounds that the self energy
function is not a gauge invariant function, and nor are the functions
appearing in the kernel; in particular, the gauge dependence of the
functions appearing in the kernel could in principle conspire to
cancel the overall factor of $(1\!-\!\xi)$ and leave a gauge
independent integral equation, which could have finite solutions.  To
answer this question it is convenient to divide the integral equation
for $\Sigma$ through by $Z(p)$ to obtain
\begin{equation} \label{n=2masseq}
M(p) = \frac{{\bar{\alpha}}}{N\pi^3} (1\!-\!\xi) \int dq \;
\frac{1}{Z(p)} \frac{M(q)}{q^2+M^2(q)} \left[ \frac{2q^2
    p}{p^2-q^2} \theta(p-q) + \frac{2 p^2 q}{q^2-p^2} \theta(q-p) \right].
\end{equation}
In the limit $p\rightarrow 0$ the function $M(p)$ is gauge invariant;
since we have shown that there can be no factor of $(1\!-\!\xi)$ from
the wavefunction renormalization factor in (\ref{n=2masseq}), it seems
that the function $M(p)$, even if the result of integrating the kernel
were finite, would retain the overall factor of $(1\!-\!\xi)$, and
therefore the only self consistent solution would be that the mass
vanished.  We have not probed the structure of the vertex $G_3$ coming
from the U(1) Ward identity \cite{clark+love}, but this will only
relate $G_3$ to combinations of $\Sigma$ and $Z$: it will not lead to
the gauge dependence required to cancel the overall factor of
$(1\!-\!\xi)$.  In order to make this argument, we depend on the
approximate form of the wavefunction renormalization
(\ref{n=2critwf}), and hence on the scale $\bar{\alpha}$; therefore
our argument does not extend back to the four dimensional model where
the scale $\bar{\alpha}$ is absent, and the coupling remains
dimensionless.

A related point of interest is that, contrasting the ${\mathcal N}=1$
case, there is no part of the integral equation for $M(p)$ which is
not troubled by the infra red divergences generated by the gauge
artifact; these divergences are known to disappear in supersymmetric
Feynman gauge.  The structure of the divergent terms make it
impossible to construct an equivalent differential equation from the
integral equation above.

\subsection{Concluding Remarks.}

In this part of the paper we have attempted to probe carefully the
gauge dependence of the ${\mathcal N}=2$ model.  We have chosen a
simplified three--vertex and we have truncated the infinite series of
higher--point vertices at four.  Within these approximations, we have
been able to compute the wavefunction renormalization, and have shown
that although the computation is plagued with (expected) infra red
divergences, we find no evidence for the overall gauge dependence
which would be necessary to cancel the factor of $(1\!-\!\xi)$ in
equation (\ref{n=2masseq}) and evade the gauge dependence argument of
reference \cite{clark+love}.  We have followed this argument through
for the (compactified) three dimensional model, where the scale
$\bar{\alpha}$ is available; in particular, the renormalization group
resummed form for the wavefunction renormalization requires the
existence of the scale $\bar{\alpha}$.  While the properties of the
four dimensional model are communicated to the three dimensional
extended model through the compactification, it is not clear that the
results presented here are applicable also for the four dimensional
model where the scale $\bar{\alpha}$ is necessarily absent; whence the
arguments of references \cite{kaiselip,appelquist98} on evading the
non--renormalization theorem in the four dimensional model could still
hold.  Our results here are in agreement with a numerical evaluation
of the effective potential for the three dimensional model in
component formalism \cite{walker99_1}.

\section{Discussion And Conclusion.}
\label{sec:discuss}

\subsection{The ${\mathcal N}=1$ Model.}

The superfield formalism we have adopted has the advantage of keeping
supersymmetry manifest; this has allowed us to concentrate on two
aspects of the infra red physics which have troubled components
computations in both the supersymmetric and non--supersymmetric
versions of this model, namely gauge invariance and the full vertex.
In particular, the superfield formalism gave us extra power to deal
with the U(1) Ward identity for the full vertex.  The disadvantages of
the superfield approach are apparent in the spurious singular terms
which appear as a result of propagating extra degrees of freedom in
the gauge superfield.  This problem can be avoided in supersymmetric
Feynman gauge, and it is this which means we have had to give separate
treatments to probe the gauge dependence and the issue of the full
vertex.

The first treatment employed some of the simplest approximations: a
near trivial vertex (which satisfies the U(1) Ward identity only in
the limit of vanishing transferred momentum, but which allows for a
non--trivial wavefunction renormalization) and neglecting the mass in
the denominators of the kernels.  This programme of approximations is
drastic, but computationally extremely convenient: enough to allow us
to probe the gauge dependence of the mass function.  This vertex
choice and the bifurcation method have been used with some success in
the study of non--supersymmetric QED${}_3$, and is sufficient to
demonstrate the existence of a dynamically generated mass
\cite{kondo+nak92} (at least in Landau gauge).  In the model studied
here, the method returns the expected critical behaviour in the
wavefunction renormalization, and exhibits a gauge independent
solution for the dynamically generated mass.  The critical behaviour
we find contrasts the trivial wavefunction renormalization of
reference \cite{koopmans89}, and is qualitatively the same as that of
non--supersymmetric QED${}_3$, where mass is also dynamically
generated.  Our second treatment employs the biggest advantage of the
superfield formalism, that of making the U(1) Ward identity tractable.
Unfortunately, the problem of infra red divergences in the superspace
approach restrict us to working in supersymmetric Feynman gauge for
this computation.  However, we find again the expected critical
behaviour in the wavefunction renormalization and again exhibit the
possibility of a finite dynamically generated mass; in particular, we
find that the solution obtained in the first simplified approach is
consistent with the full vertex, and a non--vanishing mass function in
the denominators.  It is possible to criticize the approaches taken in
both treatments we have presented, but since they tackle different
issues with the same result, we argue that taken together they provide
strong evidence for non--trivial infra red physics in ${\mathcal N}=1$
supersymmetric U(1) gauge theory.

The question of whether or not the finite mass solution is actually
selected has been addressed: contrary to the suggestion in reference
\cite{pisarski84} and the situations in the constrained models of
references \cite{alvarez78,ciuchini95,diamandis98}, we find that no extra
information can be obtained by appealing to the supersymmetry Ward
identities.  Unfortunately this leaves us with no way to determine
whether the finite mass solution is actually selected by the vacuum of
this theory.

We have found no evidence in this model for a critical flavour number,
above which no mass generation occurs.  This is in contrast to the
non--supersymmetric case \cite{appelquist86,dorey92}.  As has been
noted already, the differential equation (\ref{naivediffeq}) has
finite solutions when the mass function is gauge invariant for all $p$
(in the range permitted by our approximations: the infra red), but no
solutions which are gauge invariant only in the limit of vanishing $p$
and gauge dependent elsewhere.  This is to be contrasted with the
non--supersymmetric case, where solutions of the latter type do appear
to exist.  The integral equation for the mass function in that case
(in general gauge, and with the simplified vertex choice of \( G_3 =
Z(q)\)) is as follows:
\begin{equation}
  \label{n=0masseq}
  M(p) = \frac{8}{\pi^2 N} \left[ \int_0^p dq\; \left( \frac{1}{p} +
\frac{\xi\alpha}{2p^2} \right) \frac{q^2 M(q)}{q^2+M^2(q)} +
\int_p^\alpha dq\; \left( q+\frac{\xi\alpha}{2} \right)
\frac{M(q)}{q^2+ M^2(q)} \right].
\end{equation}
The conventional method for the non--supersymmetric case is to
consider Landau gauge ($\xi=0$) and construct an equivalent
differential equation \cite{dorey92}:
\begin{equation}
  \label{diffeqforNc}
  \frac{\partial}{\partial p} \left( p^2 \frac{\partial M}{\partial p}
\right) = -\frac{8}{\pi^2 N} \frac{p^2 \, M}{p^2+M^2};
\end{equation}
the solution of this differential equation takes the form of a
hypergeometric function; crucially this leads to a constraint on $N$
such that above a certain critical flavour number, no mass generation
occurs \cite{dorey92}.  Proceeding as we have done for the
supersymmetric model, one has to make three differentiations with
respect to $p$ to remove the integrations, and the linearized
(neglecting the mass in the denominators) result is as follows (in
dimensionless variables rescaled by $\alpha$):
\begin{equation}
  p^3 \, \dot{m}^{\prime\prime\prime} + 6p^2 \, \dot{m}^{\prime\prime}
+\left[ 6p + \frac{8}{N\pi^2} \left(p+\xi\right) \right]
\dot{m}^\prime + \frac{16}{N\pi^2} \dot{m} + \frac{8}{N\pi^2} m^\prime = 0.
\end{equation}
Demanding as before that $\dot{m}=0$ for all $p$, we find a constant
solution for $m$; substituting a gauge dependent power series
indicates that there is also the possibility of gauge independent
constant as $p\rightarrow 0$ but with a gauge dependent function of
$p$ away from this limit.  This latter type of solution does not exist
in the supersymmetric model.  We see then that the gauge dependence of
the two models is rather subtly different, and the fact that we find
no critical flavour number may be related to this.  In support of this
view it is instructive to attempt to construct a differential equation
similar to equation (\ref{diffeqforNc}) for the ${\mathcal N}=1$
supersymmetric model, which might then show some evidence for a
critical flavour number.  To this end we return to the second of
integral equations (\ref{naiveinitialgaps}) for the self energy, and
again perform differentiations with respect to $p$, but retaining the
mass as a constant in the denominators \cite{dorey92}:
\begin{equation}
  \frac{\partial}{\partial p} \left( p^2 \frac{\partial M(p)}{\partial p}
  \right) = \frac{2 (1\!-\!4\xi)}{N\pi^2} \frac{p^2 \, M(p)}{p^2+M^2}. 
\end{equation}
By analogy with the analysis of reference \cite{dorey92} the
``critical flavour number'' derived from the equation above would be
\[ N_c = \frac{ 8 (1\!-\!4\xi)}{\pi^2},\] which on account of its
explicit gauge dependence has no physical meaning.

An interesting comparison can be made with the model of reference
\cite{pisarski91}, where dynamical mass generation in an $N$--flavour
effective theory of fermions in three dimensions is studied using the
renormalization group and the $\epsilon$ expansion.  The effective
theory has for its degrees of freedom explicitly gauge independent
(composite operator) fields, and so the treatment is necessarily gauge
independent; in this approach also there is no critical flavour
number.  We also found no critical flavour number when we included the
effects of the full vertex (and the non--trivial wavefunction
renormalization): this is reminiscent of results in
non--supersymmetric QED${}_3$ \cite{pennington91} where it has been
argued that the critical flavour number of reference
\cite{appelquist86} was an artifact of including only a trivial
wavefunction renormalization and simplified vertex.  In both our
simple computation and the full computation, we have included the
effects of non--trivial wavefunction renormalization.

\subsection{The ${\mathcal N}=2$ Model.}

In contrast to the ${\mathcal N}=1$ model, in extended supersymmetric
electrodynamics, we find strong evidence for the non--existence of a
dynamically generated mass, based on a refinement of the gauge
dependence argument of reference \cite{clark+love}.  We have used the
same truncation in the Dyson--Schwinger equations (in keeping only
three-- and four--point vertices) as used in the four dimensional
model.  This truncation is consistent with restriction to the
Wess--Zumino gauge, which respects neither supersymmetry nor gauge
invariance.  Within this approximation we have chosen a simplified
vertex (but one which is consistent with a non--trivial wavefunction
renormalization) and we have then computed the wavefunction
renormalization.  We found no evidence for the appearance of an
overall factor of $(1\!-\!\xi)$ which, if present, could have evaded
\cite{kaiselip,appelquist98} the gauge dependence argument of
reference \cite{clark+love}.  Instead we found that the wavefunction
seems to have the same sort of critical behaviour as in the
non--supersymmetric and ${\mathcal N}=1$ cases, though this is by no
means certain on account of the presence of infra red divergences
(again from an artifact in the gauge superfield).  Our result here is
in line with reference \cite{walker99_1}, in which a numerical study
showed that a mass is not dynamically generated in the ${\mathcal
  N}=2$ model.  While we have been able to show that there is strong
evidence against the dynamical generation of mass in the three
dimensional extended supersymmetric model, our results depend
crucially on the existence of the scale $\bar{\alpha}$ (arising from
the size of the compactified dimension).  In the four dimensional
model, where this scale is absent, the arguments of
\cite{kaiselip,appelquist98} could still hold, and the
non--renormalization theorem of reference \cite{clark+love} be evaded.

\section*{Acknowledgements.}

The authors wish to thank G. Diamandis and B. Georgalas for helpful
discussions.  The work of N.E.M. is partially supported by P.P.A.R.C.
(U.K.) under an advanced fellowship.  A.C.--S. would like to thank the
Theory Division at CERN for their hospitality during the last stages
of this work, and gratefully acknowledges financial support from
P.P.A.R.C. (U.K.)  (studentship number 96314661), which has made his
visit to CERN possible.

\appendix

\section{Superfields In Three Dimensional ${\mathcal N}=1$ Superspace.}
\label{app:3d}

Spinor indices are raised and lowered by the antisymmetric metric
$C_{\alpha\beta}$, numerically equal to $\sigma_2$, such that
for any spinor $A$:
\begin{eqnarray}
  \label{SpIds}
  &&A^\alpha = C^{\alpha\beta} A_\beta, \nonumber\\
  &&A_\alpha = A^\beta C_{\beta\alpha}, \nonumber\\
  &&C_{\alpha\beta}C^{\mu\nu} = \delta_{[\alpha}^\mu \,
  \delta_{\beta]}^\nu ;\nonumber\\
  && A^2 \doteq \frac{1}{2} A^\alpha A_\alpha \equiv \frac{1}{2}
  C^{\alpha\beta} A_\beta A_\alpha .
\end{eqnarray}

The spinorial derivatives, covariant with respect to supersymmetry,
have the following explicit representation:
\begin{eqnarray}
  D_\alpha (x) &=& \partial_\alpha + i \theta^\beta
  \partial_{\alpha\beta},\nonumber\\ 
  D_\alpha (p) &=& \partial_\alpha + \theta^\beta p_{\alpha\beta},
\end{eqnarray}
(where \( \left\{ \partial_\alpha, \theta^\beta \right\}  =
\delta_\alpha^\beta \) ) and they obey the algebra
\begin{equation}
  \left\{ D_\alpha (p) , D_\beta (q) \right\} = (p+q)_{\alpha\beta}. 
\end{equation}
The following identities follow directly from the definition of $D$:
\begin{eqnarray}
  \label{DIds}
  &&D^\alpha (q) \,D^\beta (q) = q^{\alpha\beta} + C^{\beta\alpha}D^2 (q),
\nonumber\\
&&\delta^2(12) \, \delta^2(12) = 0,\nonumber\\
&&\delta^2(12) \,D^\alpha \,\delta^2(12) = 0, \nonumber\\
&&\delta^2(12) \, D^2 \,\delta^2(12) = \delta^2(12),\nonumber\\
&&D^2 (p) \, \delta^2(12) \, D^2 (q) \, \delta^2(12) = D^2 (p+q) \,
\delta^2(12).
\end{eqnarray}

Scalar superfields have the following component content: scalar,
spinor and auxiliary fields,
\begin{eqnarray}
  \phi (x) &=& \Phi(x,\theta)| \nonumber\\
  \psi_\alpha (x) &=&  D_\alpha \Phi(x,\theta)|
  \nonumber\\
  F(x) &=&  D^2 \Phi(x,\theta)|
\end{eqnarray}
where the symbol $|$ indicates evaluation at $\theta=0$.
Equivalently, the component content can be written:
\begin{equation}
  \Phi(x,\theta) = \phi(x) + \theta^\alpha \psi_\alpha (x) - \theta^2 F(x).
\end{equation}
The components of the connexion superfield are given by
\begin{eqnarray}
  u_\alpha (x) &=& \Gamma_\alpha (x,\theta) | \nonumber\\
  U (x) &=& \frac{1}{2} D^\alpha \Gamma_\alpha (x,\theta)| \nonumber\\
  V_{\alpha\beta} (x) &=& -\frac{i}{2} D_{(\alpha} \Gamma_{\beta)} (x,\theta) |
  \nonumber\\
  \lambda_\alpha (x) &=& \frac{1}{2} D^\beta D_\alpha \Gamma_\alpha
  (x,\theta) | .
\end{eqnarray}
The components $u_\alpha$ and $U$ can be gauged away in the so--called
Wess--Zumino gauge.  The remaining physical components are the normal
gauge field $V_{\alpha\beta}$ and its supersymmetric partner
$\lambda_\alpha$.  For further details, the interested reader is
referred to reference \cite{gates83:super}.

The superspace Feynman rules from the action in section \ref{sec:action}
are as follows:
\begin{figure}
\begin{center}
\begin{picture}(200,240)
\Line(0,190)(60,190)
\Vertex(30,190){4}
\Photon(30,194)(50,230){3}{3.5}
\LongArrow(5,185)(20,185)
\Text(15,177)[r]{$p$}
\LongArrow(40,185)(54,185)
\Text(45,177)[c]{$q$}
\LongArrow(28,205.445)(35.28,218.555)
\Text(27,216)[r]{$p\!-\!q$}
\Text(0,197)[l]{$\Phi^*$}
\Text(60,197)[r]{$\Phi$}
\Text(55,225)[l]{$\Gamma_\alpha$}
\Text(80,190)[l]{$\frac{e}{2} \,G_3 (p,p\!-\!q,q) \,C^{\alpha\beta}
  \,D_\beta (q)$}
\Line(0,110)(60,110)
\Vertex(30,110){4}
\Photon(30,114)(50,150){3}{3.5}
\LongArrow(5,105)(20,105)
\Text(15,97)[r]{$p$}
\LongArrow(40,105)(54,105)
\Text(45,97)[c]{$q$}
\LongArrow(28,125.445)(35.28,138.555)
\Text(27,136)[r]{$p\!-\!q$}
\Text(0,117)[l]{$\Phi$}
\Text(59,117)[]{$\Phi^*$}
\Text(55,145)[l]{$\Gamma_\alpha$}
\Text(80,110)[l]{$-\frac{e}{2}\,G_3 (p,p\!-\!q,q)\,C^{\alpha\beta}
  \,D_\beta (q)$}
\Line(0,30)(60,30)
\Vertex(30,30){4}
\Photon(33,32)(53,68){-3}{3.5}
\Photon(27,32)(7,68){3}{3.5}
\LongArrow(5,25)(20,25)
\Text(15,17)[r]{$p$}
\LongArrow(40,25)(54,25)
\Text(45,17)[c]{$q$}
\Text(0,37)[l]{$\Phi$}
\Text(59,37)[]{$\Phi^*$}
\Text(58,63)[l]{$\Gamma_\beta$}
\Text(2,63)[r]{$\Gamma_\alpha$}
\Text(80,30)[l]{$-\frac{ie^2}{2} \, C^{\alpha\beta}$}
\end{picture}
\caption{Interaction vertices in ${\mathcal N}=1$ model.}
\label{fig:n=1feyn}
\end{center}
\end{figure}

\section{Superfields In Four Dimensional ${\mathcal N}=1$ Superspace.}
\label{app:4d}

Spinor indices are raised and lowered by the real metrics
$\varepsilon_{\alpha\beta}$ and
$\varepsilon_{\dot{\alpha}\dot{\beta}}$, both numerically equal to the
matrix $-i \sigma_2$, such that for spinors $\chi$ and $\eta$:
\begin{eqnarray}
  \label{4dIds}
\chi^\alpha = \varepsilon^{\alpha\beta} \chi_\beta &\qquad&
\bar{\chi}_{\dot{\alpha}} = \varepsilon_{\dot{\alpha}\dot{\beta}}
\bar{\chi}^{\dot{\beta}} \nonumber\\ 
\chi\eta \doteq \chi^\alpha \eta_\alpha &\qquad& \bar{\chi}\bar{\eta}
\doteq \bar{\chi}_{\dot{\alpha}} \bar{\eta}^{\dot{\alpha}}.
\end{eqnarray}

The spinorial derivatives have the explicit representation
\begin{eqnarray}
D_\alpha &=& \partial_\alpha + i \left( \sigma^k \bar{\theta}
\right)_\alpha \partial_k,  \qquad \sigma^k \rightarrow \left
  ( {\bf{1}}, \vec{\sigma} \right), \nonumber\\
\bar{D}_{\dot{\alpha}} &=& - \bar{\partial}_{\dot{\alpha}} - i \left( \theta
  \sigma^k \right)_{\dot{\alpha}} \partial_k ,
\end{eqnarray}
where \( \{ \partial_\alpha, \theta^\beta \} = \delta_\alpha^\beta\)
and \( \{ \bar{\partial}_{\dot{\alpha}}, \bar{\theta}^{\dot{\beta}} \}
= \delta_{\dot{\alpha}}^{\dot{\beta}} \).  The covariant derivatives
obey the algebra
\begin{equation}
  \left\{ D_\alpha (p) , \bar{D}_{\dot{\alpha}} (q) \right\} = \left
  ( \sigma^k \right)_{\alpha\dot{\alpha}} \left( p+q \right)_k .
\end{equation}
The following identities follow directly from their definition:
\begin{eqnarray}
  \label{n=2DIds}
  &&\delta^2 (\bar{1}\bar{2} ) \delta^4 (12) = \delta^2 (12) \delta^4
  (12) = 0\nonumber\\
  &&D^2(q) \delta^4 (12) = -4 e^{(\theta_1-\theta_2) \slq
  \bar{\theta}_1} \delta^2 (\bar{1}\bar{2}),\nonumber\\
  &&\bar{D}^2 (q) \delta^4 (12) = -4 e^{ \theta_1 \slq
  ( \bar{\theta}_1 -\bar{\theta}_2)} \delta^2 (\theta_1 - \theta_2).
\end{eqnarray}

Chiral superfields have the following component content (where now the
symbol $|$ now indicates evaluation at $\theta=\bar{\theta}=0$):
\begin{eqnarray}
  \phi(x) &=& \Phi(x,\theta,\bar{\theta})|\nonumber\\
  \surd 2 \psi_\alpha (x) &=& D_\alpha \Phi(x,\theta,\bar{\theta}) |
  \nonumber\\
  F(x) &=& -\frac{1}{4} D^2 \Phi(x,\theta,\bar{\theta}) |,
\end{eqnarray}
which is equivalent to
\begin{eqnarray}
  \Phi(x,\theta,\bar{\theta}) &=& e^{-i \theta \not{\,\partial}
  \bar{\theta}} \Psi(x,\theta) \nonumber\\
  \Psi(x,\theta) &=& \phi(x) + \surd 2 \theta \psi(x) + \theta^2 F(x).
\end{eqnarray}
The component content for an antichiral superfield is similar, with
$\theta$ replaced by $\bar{\theta}$, and $\bar{D}$ replacing $D$.  The
component content of the gauge superfield is as follows:
\begin{eqnarray}
  &&C(x) = V(x,\theta,\bar{\theta}) | \nonumber\\
  &&\chi_\alpha(x) = -i D_\alpha V(x,\theta,\bar{\theta}) | \qquad
  \bar{\chi}_{\dot{\alpha}} (x) = \bar{D}_{\dot{\alpha}}
  V(x,\theta,\bar{\theta}) | \nonumber\\
  &&\left( M(x)+iN(x) \right) = \frac{i}{2} D^2
  V(x,\theta,\bar{\theta}) | \qquad \left( M(x) - i N(x) \right) =
  -\frac{i}{2} \bar{D}^2 V(x,\theta,\bar{\theta}) | \nonumber\\
  &&\lambda_\alpha (x) = -\frac{i}{4} \bar{D}^2 D_\alpha
  V(x,\theta,\bar{\theta}) | \qquad \bar{\lambda}_{\dot{\alpha}} (x) = \frac{i}{4} D^2
  \bar{D}_{\dot{\alpha}} V(x,\theta,\bar{\theta}) | \nonumber\\
  &&\left( \sigma^k \right)_{\alpha\dot{\alpha}}  v_k (x) = - \frac{1}{2} \left
  [ D_\alpha,\bar{D}_{\dot{\alpha}} \right] V(x,\theta,\bar{\theta}) |
  \nonumber\\
  && d(x) = \frac{1}{8} D^\alpha \bar{D}^2 D_\alpha
  V(x,\theta,\bar{\theta}) | \nonumber\\
  && 4 \delta_\alpha^\beta d(x) - 2i \left( \sigma^m
  \bar{\sigma}^n \right)_\alpha^\beta \partial_{[m} v_{n]} (x)  =
  D^\beta \bar{D}^2 D_\beta V(x,\theta,\bar{\theta}) | .
\end{eqnarray}
In Wess--Zumino gauge the components $C$, $\chi$, $\bar{\chi}$, $M$
and $N$ can all be gauged away, leaving the component expansion
\begin{eqnarray}
  V (x,\theta,\bar{\theta}) &=& - \left( \theta \sigma^k \bar{\theta}
  \right) v_k (x)  + i \theta^2 
  \bar{\theta} \bar{\lambda}(x) - \bar{\theta}^2 \theta \lambda (x) +
  \frac{1}{2} \theta^2 \bar{\theta}^2 d(x), \nonumber\\
  V^2 (x,\theta,\bar{\theta}) &=& -\frac{1}{2} \theta^2 \bar{\theta}^2
  v^2 (x), \nonumber\\ 
  V^3 (x,\theta,\bar{\theta}) &=& 0.
\end{eqnarray}
For further details the interested reader is referred to
\cite{srivastava}.  The Feynman rules are very simple; outside the
Wess--Zumino gauge there are an infinite number of possible vertices,
with two external matter superfield lines (one chiral and one
antichiral), and $n$ gauge superfields, the vertex corresponding to a
factor of $i(2e)^n$ (or $i(-2e)^n$ for the fields $\Phi_-$ and
$\Phi_-^\dagger$, which are negatively charged under U(1)).  In
Wess--Zumino gauge, the series terminates at $n=2$, and we dress the
corresponding vertices with the functions $G_3$ and $G_4$.



\end{document}